\documentclass[10pt,aps,prb,twocolumn,showpacs,superscriptaddress]{revtex4-1}

\usepackage{graphicx}  
\usepackage{dcolumn}   
\usepackage{bm}        
\usepackage{amsmath}   
\usepackage{amssymb}   

\newcommand{\Qi}{\mathcal{Q}i} 
\newcommand{\n}[1]{\hat{n}_{#1}} 
\newcommand{\create}[2]{\hat{#1}^{\dagger}_{#2}} 
\newcommand{\annihilate}[2]{\hat{#1}^{\phantom{\dagger}}_{#2}} 
\newcommand{\ket}[1]{\left|#1\right\rangle} 
\newcommand{\bra}[1]{\left\langle#1 \right|} 

\newcommand{\trans}[1]{\mathcal{U} #1 \mathcal{U}^\dagger}
\newcommand{\vS}{\vec{S}}
\newcommand{\vPauli}{\vec{\tau}}

\usepackage[usenames]{xcolor}

\begin{document}
\title[Quantum Interference in Off-Resonant Transport]
{Quantum Interference in Off-Resonant Transport through Single Molecules}

\author{Kim G.~L.~Pedersen}
\email{kim@fys.ku.dk}
\affiliation{Nano-Science Center, University of Copenhagen}
\affiliation{Niels Bohr Institute, University of Copenhagen}
\author{Mikkel Strange}
\affiliation{Nano-Science Center, University of Copenhagen}
\affiliation{Department of Chemistry, University of Copenhagen}
\author{Martin Leijnse}
\affiliation{Solid State Physics and Nanometer Structure Consortium (nmC@LU), Lund University, 221 00 Lund, Sweden}
\author{Per Hedeg{\aa}rd}
\affiliation{Nano-Science Center, University of Copenhagen}
\affiliation{Niels Bohr Institute, University of Copenhagen}
\author{Gemma C. Solomon}
\affiliation{Nano-Science Center, University of Copenhagen}
\affiliation{Department of Chemistry, University of Copenhagen}
\author{Jens Paaske}
\affiliation{Nano-Science Center, University of Copenhagen}
\affiliation{Niels Bohr Institute, University of Copenhagen}
\affiliation{Center for Quantum Devices, Niels Bohr Institute, University of Copenhagen}

\date{\today}

\begin{abstract}

\noindent
We provide a simple set of rules for predicting interference effects in off-resonant transport through single-molecule junctions. These effects fall in two classes, showing respectively an \textit{odd} or an \textit{even} number of nodes in the linear conductance within a given molecular charge state, and we demonstrate how to decide the interference class directly from the contacting geometry. For neutral alternant hydrocarbons, we employ the Coulson-Rushbrooke-McLachlan pairing theorem to show that the interference class is decided simply by tunneling on and off the molecule from same, or different sublattices. More generally, we investigate a range of smaller molecules by means of exact diagonalization combined with a perturbative treatment of the molecule-lead tunnel coupling. While these results generally agree well with GW calculations, they are shown to be at odds with simpler mean-field treatments. For molecules with spin-degenerate ground states, we show that for most junctions, interference causes no transmission nodes, but argue that it may lead to a non-standard gate-dependence of the zero-bias Kondo resonance.

\end{abstract}

\maketitle


\section{Introduction}

\begin{figure}[tb]
  \begin{center}
    \includegraphics[width=.8\columnwidth]{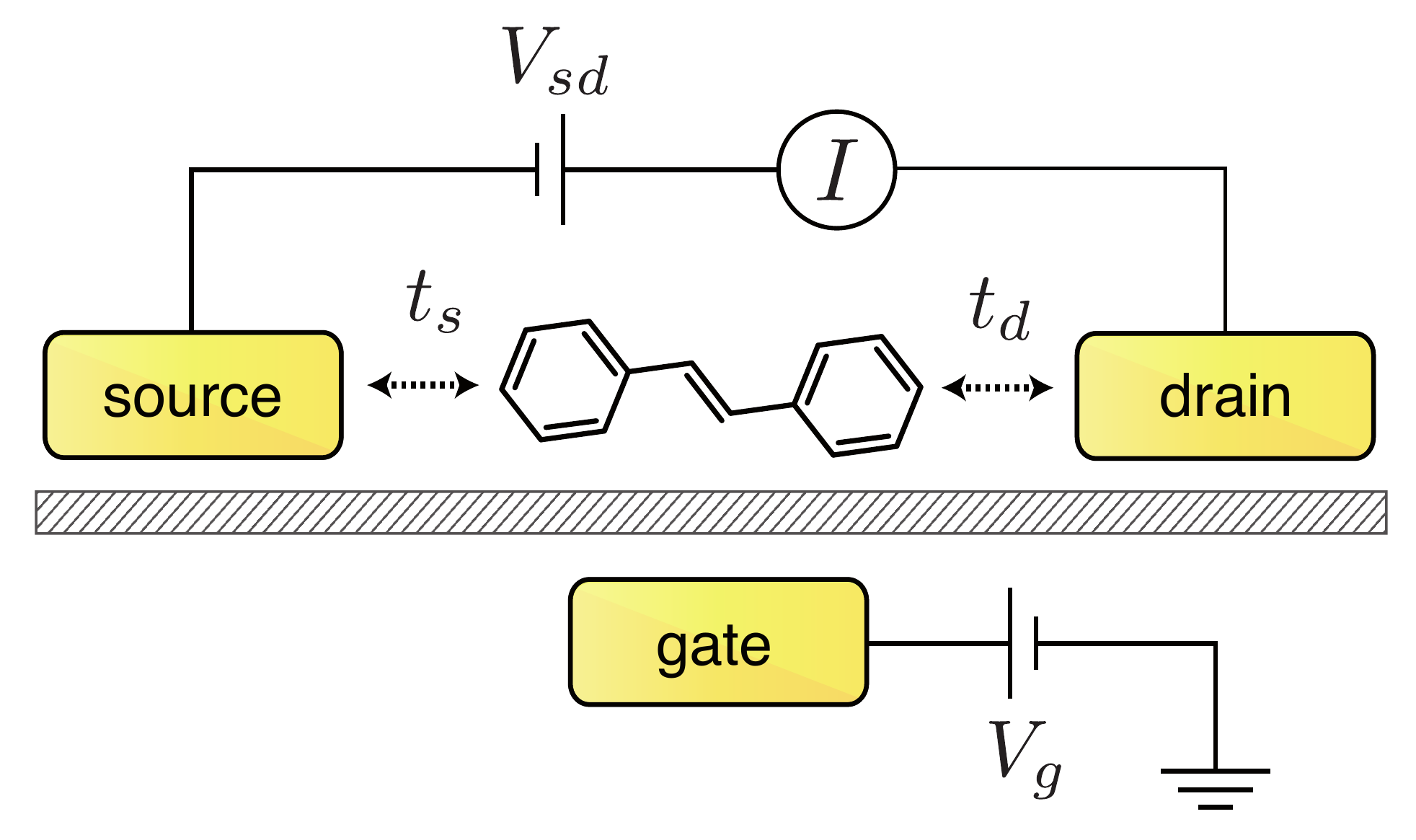}
  \end{center}
  \caption{(color online) The molecular junctions setup with a source-drain voltage $V_{sd}$ applied across the molecule, which is tunnel-coupled to leads through $t_{s}$ and $t_{d}$. The electrostatic environment is controlled through a backgate voltage $V_g$.}
  \label{fig:model}
\end{figure}

Interference effects in electronic transport through single molecule junctions have recently attracted a lot of attention as they offer a sensitive handle with which to tune transport properties~\cite{Baer2002,VanDijk2006,Duchemin2008,Karlstrom2011,Markussen2010,Bergfield2010,Arroyo2013}. Single molecule junctions and quantum dots exhibit many similar transport features, including low-temperature observations of Coulomb blockade and possibly Kondo effect for spin-degenerate molecules, whenever a backgate (cf. Fig.~\ref{fig:model}) has been available for tuning the molecular energy levels.

For quantum dots, interference effects have been studied intensively in the context of the phase lapses, which were observed, by implementing the dot into one arm of an Aharonov-Bohm interferometer ~\cite{Yacoby1995,Schuster1997}. In this context, a number of theoretical works have investigated the possibility for interference induced transmission zeros in quantum dots of various shapes and sizes, possibly involving multiple connections to the leads, possibly including effects of disorder and/or interactions~\cite{Lee1999,Silvestrov2000,Yeyati2000,Hackenbroich2001,Silva2002,Molina2013,Karrasch2007,Hecht2009}. Whereas all of these different factors have been shown to play a determining role for interference effects in quantum dots, the largely random element of dot shape alone makes these effects more or less serendipitous and difficult to employ in an intentional design. This problem is bypassed when replacing the dot by a single molecule, where the interference is controlled by the well-defined electronic and magnetic structure prescribed by the chemical synthesis.

Transport through a single molecule junction can be on- or off-resonant, depending on the relative strengths of molecule-lead couplings to the addition energy of the molecule, as well as on the position of the energy-levels of the molecule relative to the chemical potentials of the metallic leads~\cite{Moth2009}.
In this paper, we deal exclusively with molecules out of resonance, where it makes sense to talk about a ground state for the junction having a definite number of electrons on the molecule. This is the, already well-documented~\cite{Kubatkin2003,Osorio2007,Roch2008,Osorio2008,Osorio2010,Osorio2010b,Lee2011,Prins2011,Fock2012}, regime in which three-terminal experiments observe Coulomb diamonds (cf. Fig.~\ref{fig:stilbene_transmission}(b)). In the low-conductance interior of these diamonds, transport takes place via virtual charge fluctuations of the molecule, and safely inside a diamond this so-called cotunneling conductance can be reliably calculated by perturbation theory in the weak charge fluctuations, thus relying on the effective lead-molecule coupling being much smaller than the addition energy.

In practice, even physical phenomena like the Kondo effect, involving 'not so weak' virtual charge fluctuations, are well described within a simple cotunneling model~\cite{Scott2010}. Calculations may require non-perturbative methods, but the effect itself requires nothing but repeated cotunneling processes. This line of reasoning works very well even for quantum dots with addition energies of only a few meV and a lack off good separation of energy scales~\cite{Kouwenhoven2001,Holm2008}. Even quantitative descriptions of line shapes in inelastic cotunneling spectroscopy can be carried out using a simple cotunneling model as the starting point for transport calculations~\cite{Paaske2006}. In comparison, typical molecules under consideration easily exhibit addition energies of the order of 100 meV, with huge Coulomb diamonds ensuring a much better separation of energy scales. This is what makes a gated single molecule junction such an exceptional system for high-quality inelastic cotunneling spectroscopy, resolving magnetic, or vibrational excitations on the scale of a few meV~\cite{Roch2008,Osorio2008,Osorio2010,Fock2012}. Unless the addition energy somehow becomes smaller than the tunnel broadening, an effective cotunneling model thus provides a very simple point of departure for a perturbative treatment of charge fluctuations, which remains valid throughout the parameter space of backgate, $V_g$, and bias voltage, $V_{sd}$, except for the crossing lines of mixed valance or charge-degeneracy. Only on these tunnel-broadened lines the cotunneling model breaks down, as the molecule is tuned into resonance.

In this off-resonant regime, electrons are not streaming through different arms of a coherent wave-guide, but rather traversing the molecule by means of coherent cotunneling processes involving the virtual tunneling of either an electron or a hole. The aim of this paper is to sort out the basic mechanisms for interference in this cotunneling regime, and to provide a simple and robust means of categorizing the possible interferences to be expected for a given molecule. What interference effects may be expected in this regime, and to what extent might one already capture these within a simple mean-field treatment of the interacting molecular $\pi$-electron system?
We work out a simple set of rules for predicting interference effects in off-resonant transport through alternant (bipartite) hydrocarbons based on the contacting geometry alone. The rules are based on the Coulson-Rushbrooke-McLachlan pairing theorem~\cite{Coulson1940,McLachlan1959,McLachlan1961}, from which we derive a relation between tunneling amplitudes for respectively holes, and electrons, which constitute the two interfering amplitudes in a cotunnel-junctions. For a single orbital model, we simply rederive the well-known impurity physics fact, that the Anderson model has no potential scattering term at the particle-hole symmetric point. For a real molecule, however, we uncover a set of non-trivial, yet easy to use, graphical rules for deciding if the junction will show an even or an odd number of interference dips in the zero-bias conductance, as the gate-voltage is varied across the relevant charge-state of the molecule.

To support our findings we investigate the effect numerically. For sufficiently small molecules, we perform an exact diagonalization (ED) of the interacting molecular $\pi$-system, and demonstrate how this rephrased theorem works for strictly off-resonant transport. For a number of different molecules, we compare the result with those obtained using other popular methods like density functional theory (DFT), H\"{u}ckel theory (HT), and GW, which all include higher order tunneling (hybridization) effects but treat the interactions only approximately. For most situations even a simpel HT is shown to predict the \textit{correct interference class}, but most often the effective single-particle calculations (HT, DFT, HF) return the \textit{incorrect dip positions} and even spurious dip-degeneracies, which are usually lifted by GW and ED calculations. Our simple classification rules provide a valuable tool for gauging the validity of approximate calculations, and since the rephrased pairing theorem is topological in nature, we expect it to be of more general validity beyond the restricted class of neutral homo-atomic alternant hydro-carbons.


\section{The PPP Model for conjugated molecules}

In a molecular junction the molecule is tunnel-coupled to two electrodes at a bias voltage $V_{sd}$, and with a backgate voltage $V_g$ controlling the electrostatic environment (see Fig.~\ref{fig:model}). The molecular $\pi$-system is modeled by the semi-empirical Pariser-Parr-Pople model~\cite{Pople1953a,Pariser1953}:
\begin{align}
\hat{H}&=
  \sum_{\langle i,j\rangle} \sum_{\sigma=\uparrow/\downarrow} \left(t_{ij} \create{c}{i\sigma} \annihilate{c}{j\sigma} +  h.c. \right)- e V_g \sum_i (\n{i}-1) \nonumber\\
  &+\sum_i U (\n{i \uparrow}-\tfrac{1}{2})(\n{i \downarrow}- \tfrac{1}{2})
  +\frac{1}{2} \sum_{i\neq j} V_{ij} (\n{i} - 1)(\n{j} -1).\nonumber
\end{align}
The operator $\create{c}{i \sigma}$ creates an electron with spin $\sigma$ on the $p_z$-orbital $|i\rangle$, $\hat{n}_{i \sigma} = \create{c}{i \sigma} \annihilate{c}{i \sigma}$ and $\hat{n}_{i} = \n{i \uparrow} + \n{i \downarrow}$. The Coulomb interaction is given by the Ohno parametrization~\cite{Ohno1964} $V_{ij} = U/(\sqrt{1 + |\vec{r}_{ij}|^2 U^2/207.3 \text{ eV}})$, where $|\vec{r}_{ij}|$ is the real-space distance between two $p_z$-orbitals $|i\rangle$ and $|j\rangle$ measured in {\AA}ngstr\"{o}m. For $sp^2$ hybridised carbon, the nearest neighbor overlap, $t_{ij}$, is $t\approx - 2.4$ eV, and $U \approx 11.26$ eV~\cite{Soos1984}. Absorbing a constant capacitive lever arm, the backgate voltage shifting the molecular energy levels is denoted by $V_g$.

The isolated molecular $\pi$-system with $N$ electrons has eigenenergies $E^N_n$, with corresponding many-body eigenstates $\ket{\Psi^N_{n}}$. The $N$-electron excitation spectrum is given by $\varepsilon_n^0 = E_n^{N}-E_0^N$, the energy costs of adding an electron to the $\pi$-system is given by $\varepsilon_{n}^p = E_n^{N+1}-E_0^N$, and the costs of removing an electron by $\varepsilon_{n}^h = E_0^N - E_n^{N-1}$. For the molecules investigated below, these eigenenergies and many-body eigenstates are determined numerically by exact diagonalization~\cite{Siro2012}.

The source and drain ($\alpha=s,d$) electrodes are modeled by non-interacting electrons with constant densities of states $\rho_\alpha$. Since interference effects depend crucially on the entry and exit points for the transport electrons, we assume that only one $p_z$-orbital $|i_\alpha \rangle$ couples to each lead $\alpha = s,d$ and the tunneling term is included as $\hat{H}_H = \sum_{\alpha k\sigma} ( t_{\alpha} \create{c}{i_\alpha \sigma} \annihilate{c}{\alpha k \sigma} + h.c.)$. When the coupling strengths $\Gamma_\alpha = 2 \pi \rho_\alpha |t_\alpha|^2$ to source and drain electrodes are much smaller than the excitation energies $\Gamma_{s,d}\ll |\varepsilon_0^p|,|\varepsilon_0^h|$, the molecular junction is \emph{blockaded}, and transport is restricted to cotunneling processes via virtual charge-fluctuations of the molecule. The molecule holds a definite number of electrons and the off-resonant current is determined by leading order perturbation theory in $\Gamma_{s}\Gamma_{d}/(\varepsilon_{0}^{p,h})^{2}$.


\section{Transport through a spin-singlet ground state}

For a \textit{non-degenerate} molecular ground state, there are only two contributing transport processes. One process transfers an electron (\textit{p}) from source to drain, while the second process transfers a hole (\textit{h}) in the reverse direction. At zero bias voltage and safely away from the charge degeneracy points, i.e. $e V_{g}=\varepsilon_{p}^{0}, \varepsilon_{h}^{0}$, the zero-temperature off-resonant current can be calculated from the generalized Fermi golden rule as~\cite{BruusFlensberg} 
\begin{align}
  I(V_g,V_{sd})
  &=
  \frac{e}{h} \Gamma_s \Gamma_d
  \sum_{m,\sigma, \gamma}
  \int_{-e V_{sd}/2+\varepsilon_m^0}^{e V_{sd}/2}
  \mathrm{d} \omega
  \theta(|e V_{sd}| - \varepsilon_m^0 )
  \nonumber \\
  &\quad \times
   \left|
     h^{\gamma\sigma}_{m,0} (V_g,\omega-\varepsilon_m^0)+
     p^{\gamma\sigma}_{m,0} (V_g,\omega)
   \right|^2,
   \label{eq:singlet_current}
\end{align}
where $m$ runs over the $N$-particle eigenspectrum, while $\sigma, \gamma$ both run over spin $\uparrow, \downarrow$. We have also introduced the cotunneling amplitudes,
 \begin{align*}
  h^{\gamma\sigma}_{m,l}(V_g, \omega) &= \sum_{n} \frac{
  \bra{\Psi_m^{N}} \create{c}{i_s \gamma} \ket{\Psi_n^{N-1}} \bra{\Psi_n^{N-1}} \annihilate{c}{i_d \sigma} \ket{\Psi_l^{N}}
  }{e V_g - \omega - \varepsilon_n^h - i 0^+}, \\
  p^{\gamma\sigma}_{m,l} (V_g, \omega) &=  \sum_{n} \frac{
    \bra{\Psi_m^N}\annihilate{c}{i_d \sigma} \ket{\Psi_n^{N+1}} \bra{\Psi_n^{N+1}}\create{c}{i_s \gamma}\ket{\Psi_l^N}
    }{e V_g - \omega - \varepsilon_n^p + i 0^+}. 
\end{align*}
These amplitudes are given in terms of the molecular energy spectrum and the Feynman-Dyson (FD) orbitals, i.e. the correlated generalization of molecular orbitals, calculated as the matrix elements of the hole creation operator, $\bra{\Psi_n^{N-1}} \annihilate{c}{i \sigma} \ket{\Psi_m^{N}}$ and the electron creation operator $\bra{\Psi_n^{N+1}} \create{c}{i \sigma} \ket{\Psi_m^{N}}$ in the basis of the relevant many-particle molecular eigenstates.

This formula for the off-resonant current omits any bias-dependence of excited state occupations, which would play a role for the detailed shape of inelastic cotunneling steps (cf. e.g. Ref.~\onlinecite{Begemann2010}),  but is immaterial for the present discussion.
Henceforth, we shall mostly consider the zero-bias cotunnelling conductance, \mbox{$G = \lim_{V_{sd} \rightarrow 0} dI/dV_{sd}$}, which can be written on the form
\begin{align}
  G = G_0\Gamma_s \Gamma_d
  \sum_{\sigma}
   \left|
     h^{\sigma\sigma}_{0,0} (V_g,0)+
     p^{\sigma\sigma}_{0,0} (V_g,0)
   \right|^2,
   \label{eq:Gzero}
\end{align}
in terms of the quantum of conductance, $G_0 = e^2/h$. In this case, the zero-bias conductance becomes synonymous with the junction transmission, i.e. $T = G/G_0$.

\subsection{Interference classification}

\begin{figure}[t]
   \centering
     \includegraphics[width=\columnwidth]{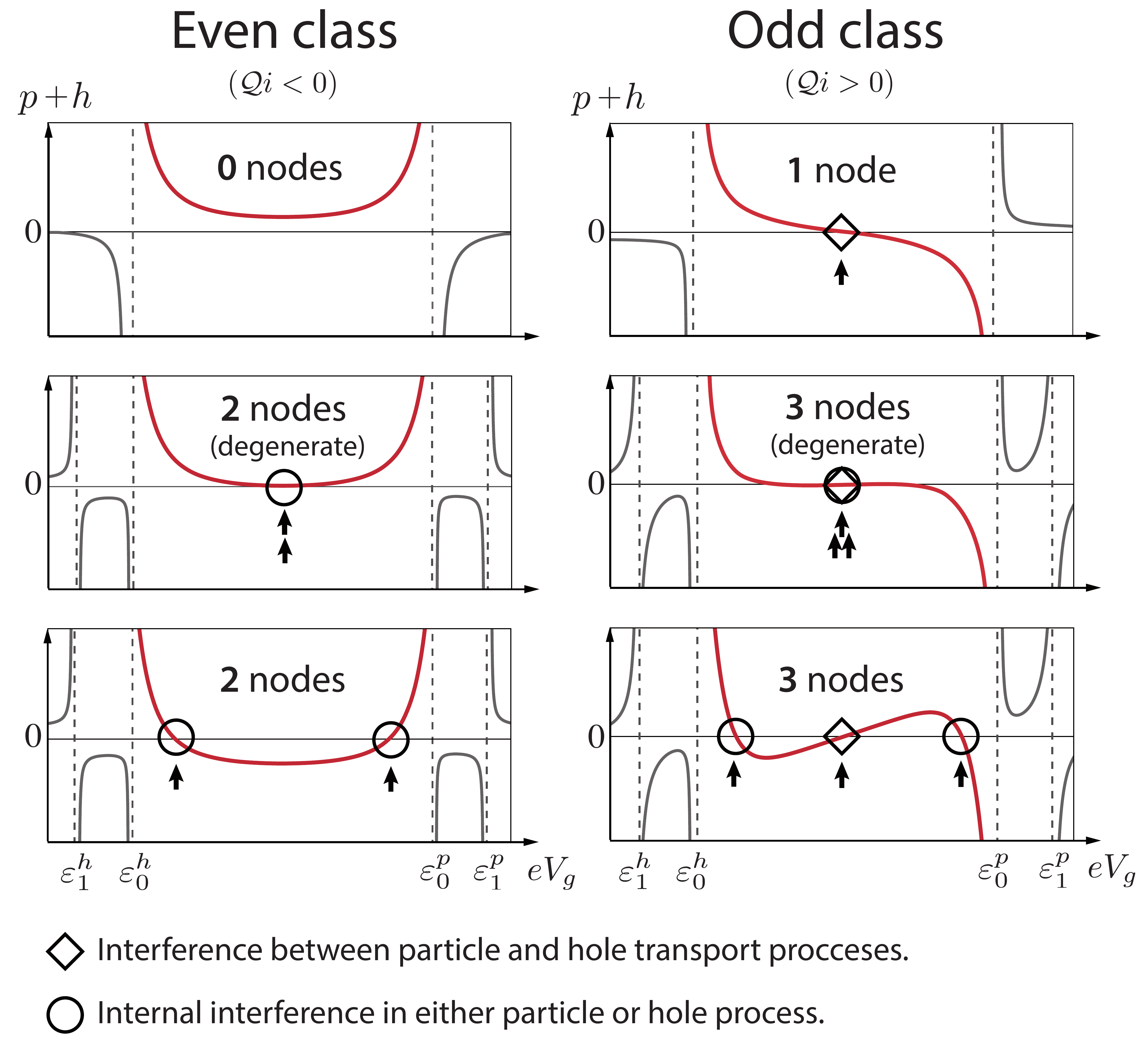}
       \caption{(color online) The even and odd quantum interference classes. The signs of the pole of $h^{\sigma \sigma}_{00}$ at $\varepsilon_0^h$ and of $p^{\sigma \sigma}_{00}$ at $\varepsilon_0^p$ determine the interference class uniquely. The red line shows the amplitude in the valid regime of $V_g$. The interference nodes can be interpreted as happening between the particle and hole processes (diamonds), or within the particle or hole processes (circles). Note the anomalous degenerate cases with only one node, which shows up in some DFT and HT calculations.}
   \label{fig:qi}
\end{figure}
Based on this zero-bias conductance formula, we divide interference effects in off-resonant molecular junctions into classes of \textit{even} or \textit{odd}, depending on the number of conductance zeros found by varying the gate voltage all the way across a given charge state. As summarized graphically in Fig.~\ref{fig:qi}, divergences of the $h_{0,0}^{\sigma \sigma}$ amplitude at $\varepsilon_0^h$ and similarly of $p_{0,0}^{\sigma \sigma}$ at $\varepsilon_0^p$ impose severe constraints on the zero bias conductance:  1) When both divergences share the same sign, one goes to infinity, while the other goes to minus infinity, when approaching the poles from a point in between $\varepsilon_0^h$  and $\varepsilon_0^p$. This forces the total conductance to have an \textit{odd} number of zeros. 2) When the divergences have opposite signs, they both go to either plus or minus infinity, which forces the total conductance to have an \textit{even} number of zeros in between. The relative sign of the divergences is completely determined by the numerators of $h^{\sigma \sigma}_{0,0}$ and $p^{\sigma \sigma}_{0,0}$, and the classification is therefore encoded in the sign of their ratio:
\begin{align}\label{eq:Qi}
  \Qi \equiv
    \frac{
    	\bra{\Psi_0^N} \create{c}{i_s \sigma} \ket{\Psi_0^{N-1}} \bra{\Psi_0^{N-1}}\annihilate{c}{i_d \sigma}\ket{\Psi_0^N}
    }{
    	\bra{\Psi_0^N}\annihilate{c}{i_d \sigma} \ket{\Psi_0^{N+1}} \bra{\Psi_0^{N+1}}\create{c}{i_s \sigma}\ket{\Psi_0^N}
    },
\end{align}
where a sum over any ground state degeneracies of the $N\pm 1$ charge states is implied. When $\Qi>0$ the numerators share the same sign and the interference class is odd, and when $\Qi<0$ the numerators have opposite signs and the interference class is even.

For a simple non-interacting (H\"{u}ckel) model of the molecule, the numerator of $h_{0,0}$ ($p_{0,0}$) is the product of the HOMO (LUMO) wavefunction on the sites connected to source and drain. For such models the relation between interference and relative sign of HOMO and LUMO has been investigated previously~\cite{Tada2002,Tada2003,Tada2004,Yoshizawa2008,Lovey2012}. As we shall demonstrate below, however, intra-molecular interactions may affect the interference nodes in the conductance.

This classification highlights the interference mechanism responsible for the various possible nodes in the conductance. When hole and particle transport amplitudes cancel, the result is exactly \textit{one} node. All remaining interference nodes of either class can be interpreted as happening completely within a hole (or a particle) transport amplitude. This is indicated in Fig.~\ref{fig:qi} where a diamond marks nodes interpreted as arising from particle-hole interference, while a circle marks nodes due interference solely within hole (or particle) processes. Note that the $\Qi$ classification parameter is readily generalized to the case when many orbitals connect to each electrode, by replacing $\bra{\Psi_m^N}\annihilate{c}{i_{\alpha} \sigma} \ket{\Psi_n^{N+1}}$ with an average over all connected orbitals $\ket{i_s}$ weighted by their coupling strengths $|t_{\alpha i_\alpha}|^2$.

\subsection{The Starring Rule for Alternant Hydrocarbons}

\begin{figure*}[tbh]
  \includegraphics[width = .32\textwidth]{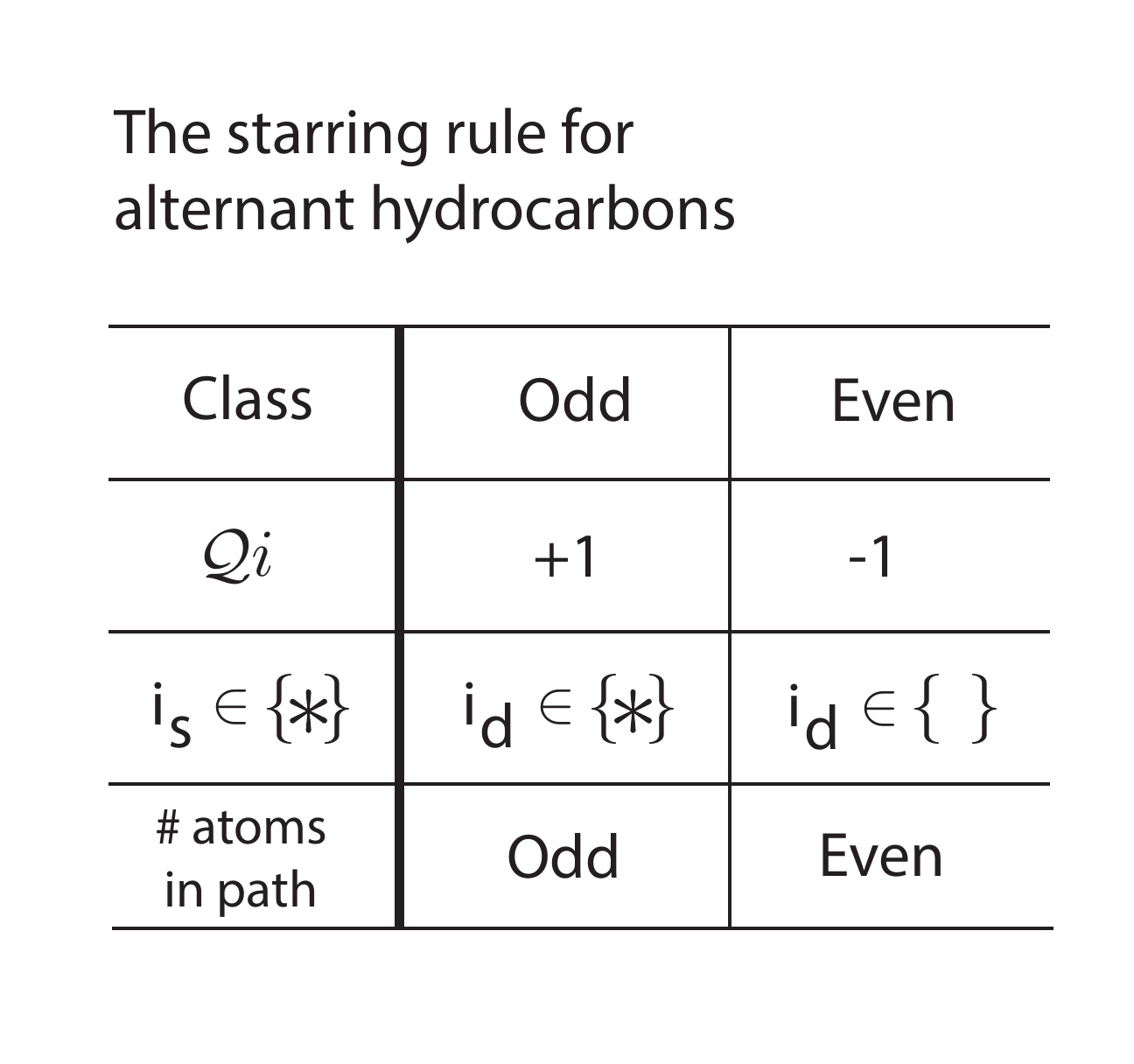}
  \includegraphics[width = .32\textwidth]{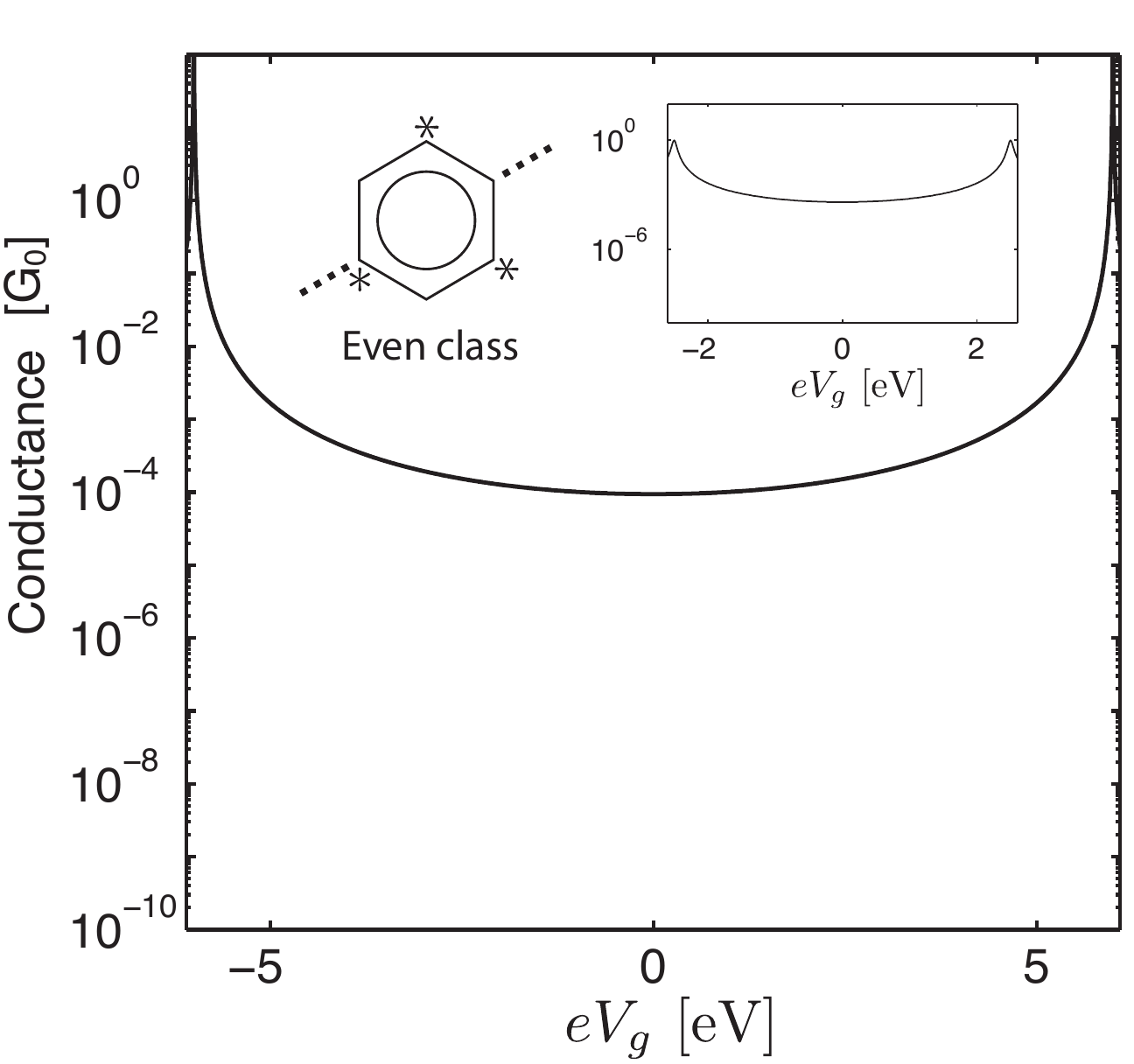}
  \includegraphics[width = .32\textwidth]{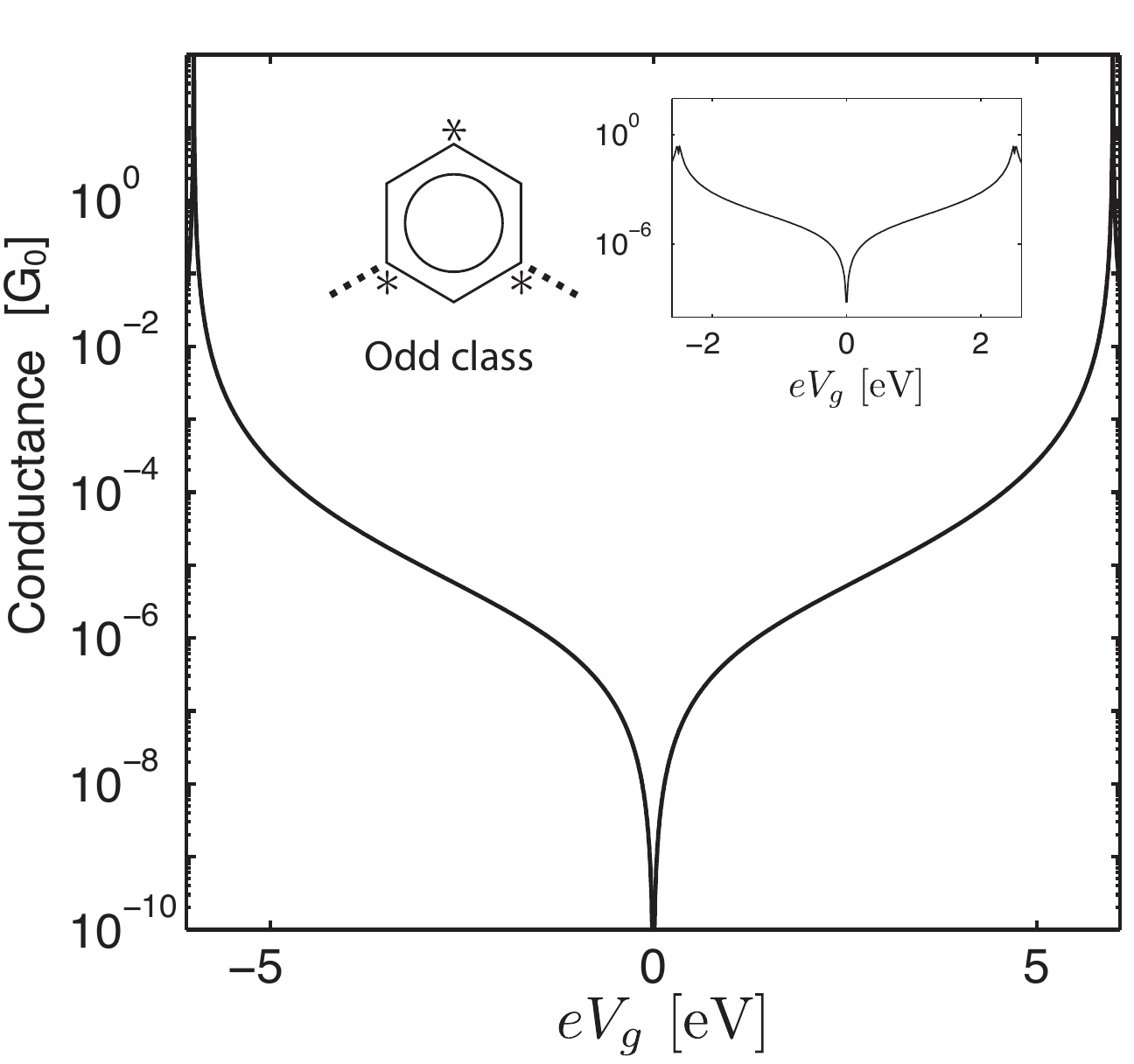}\\
  \includegraphics[width = .32\textwidth]{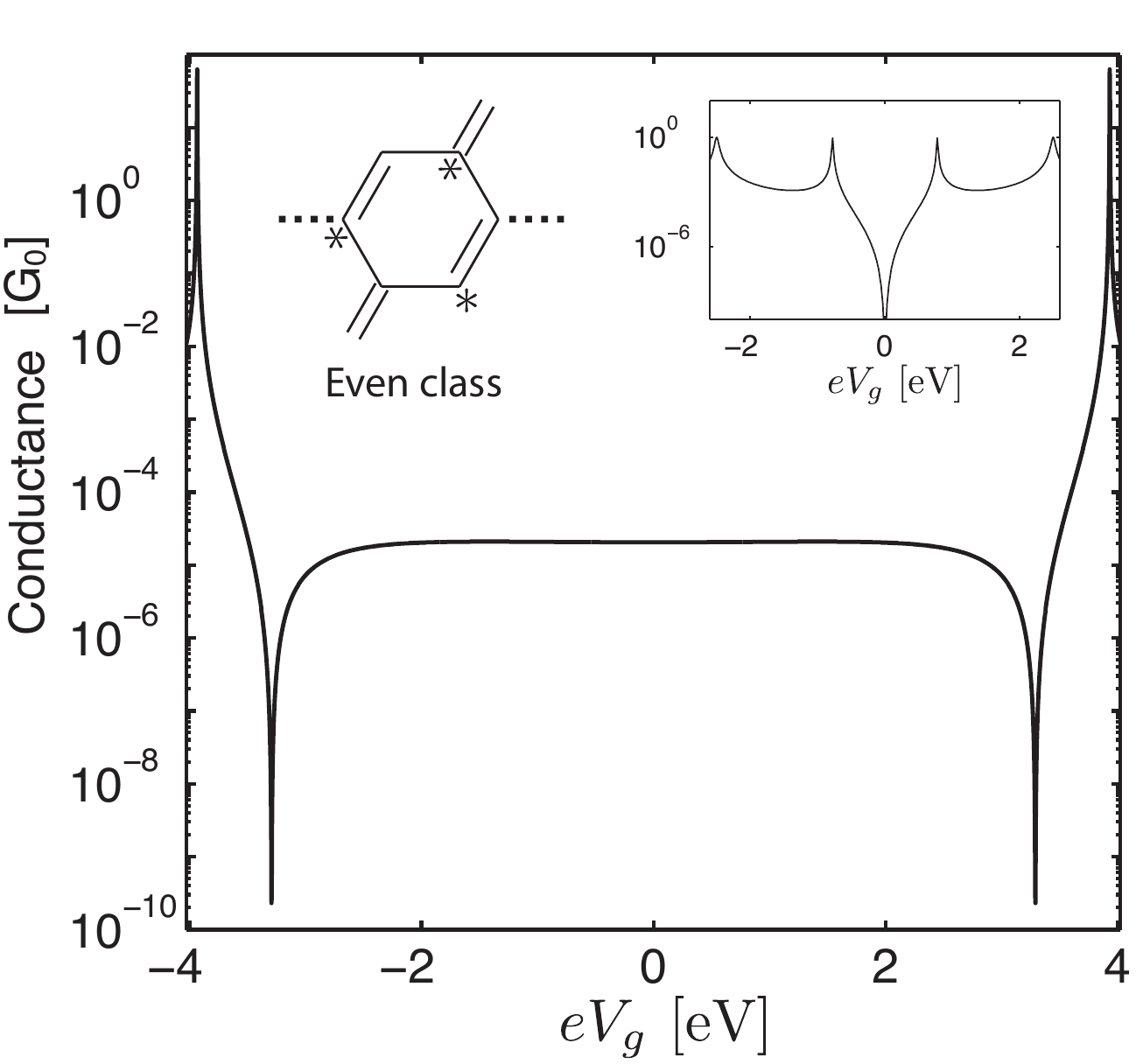}
  \includegraphics[width = .32\textwidth]{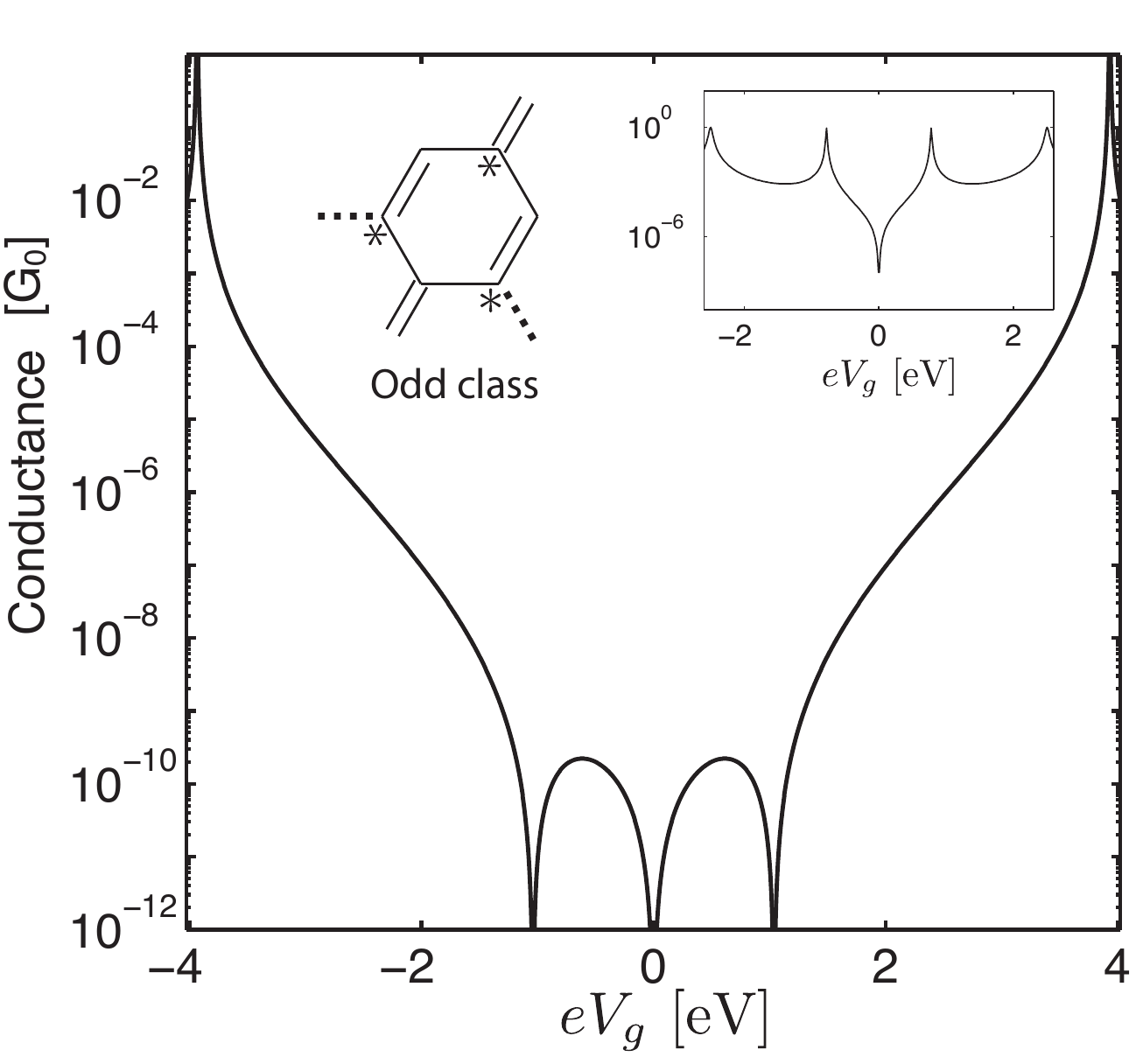}
  \includegraphics[width = .32\textwidth]{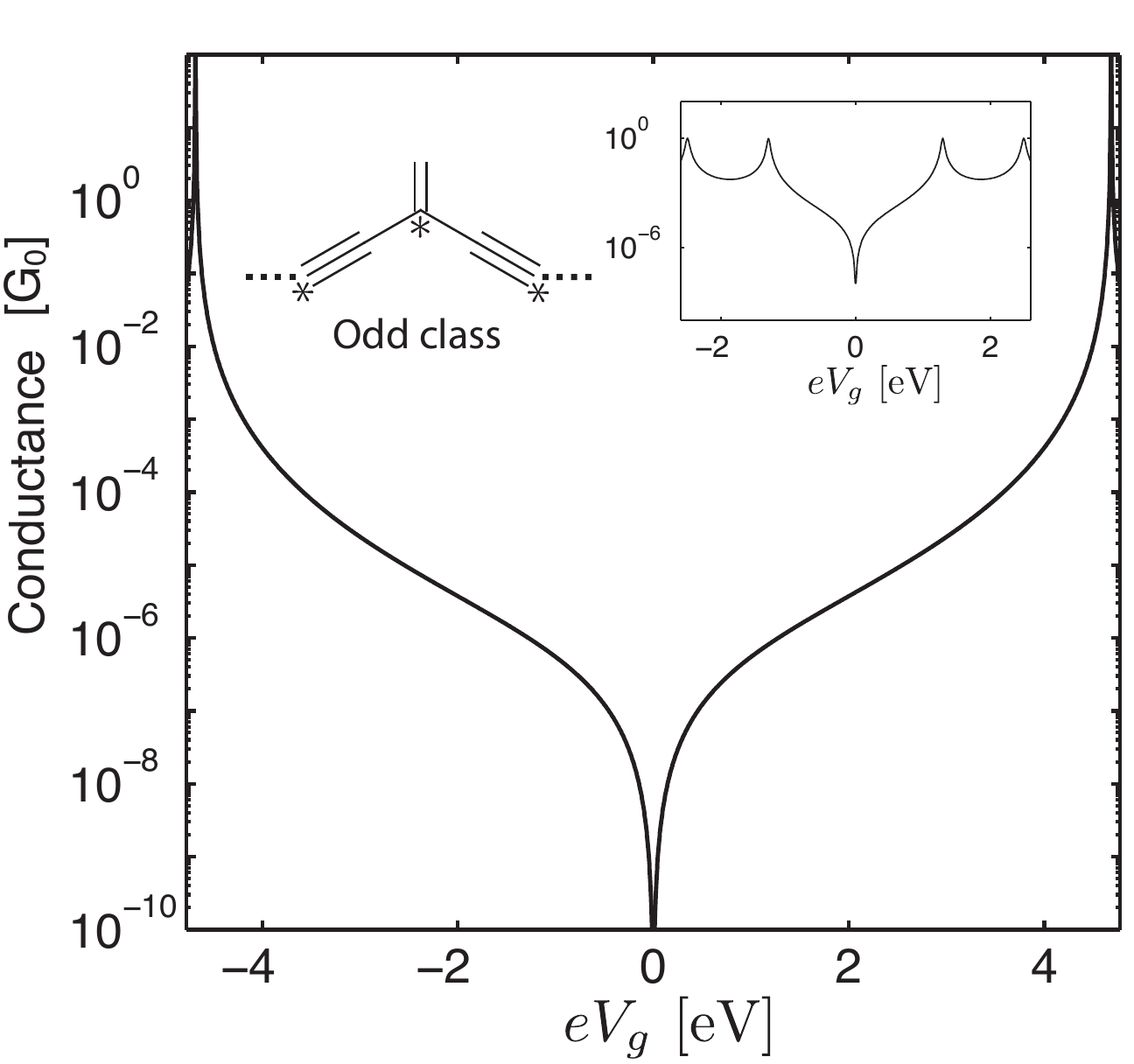}\\
  \includegraphics[width = .32\textwidth]{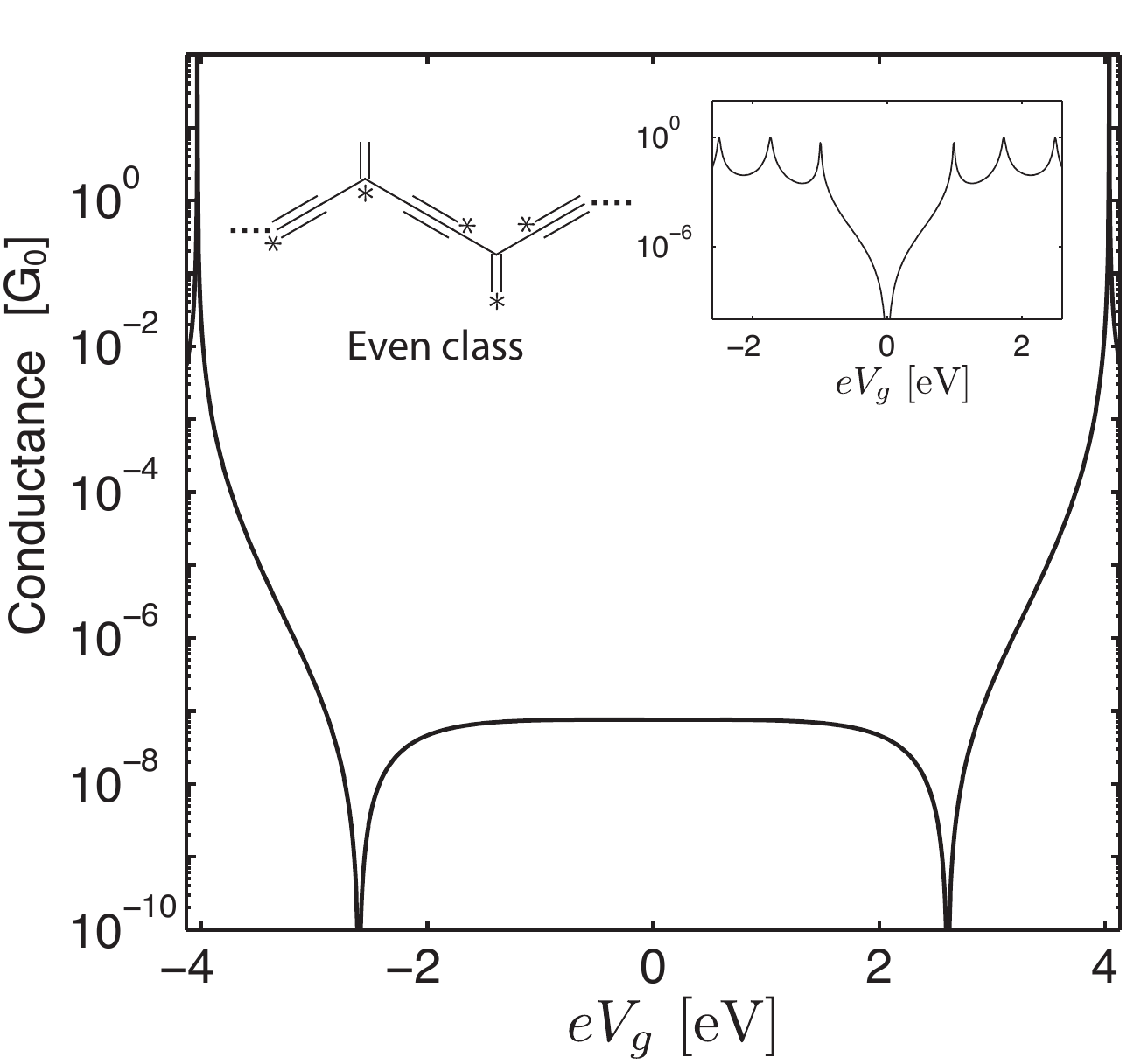}
  \includegraphics[width = .32\textwidth]{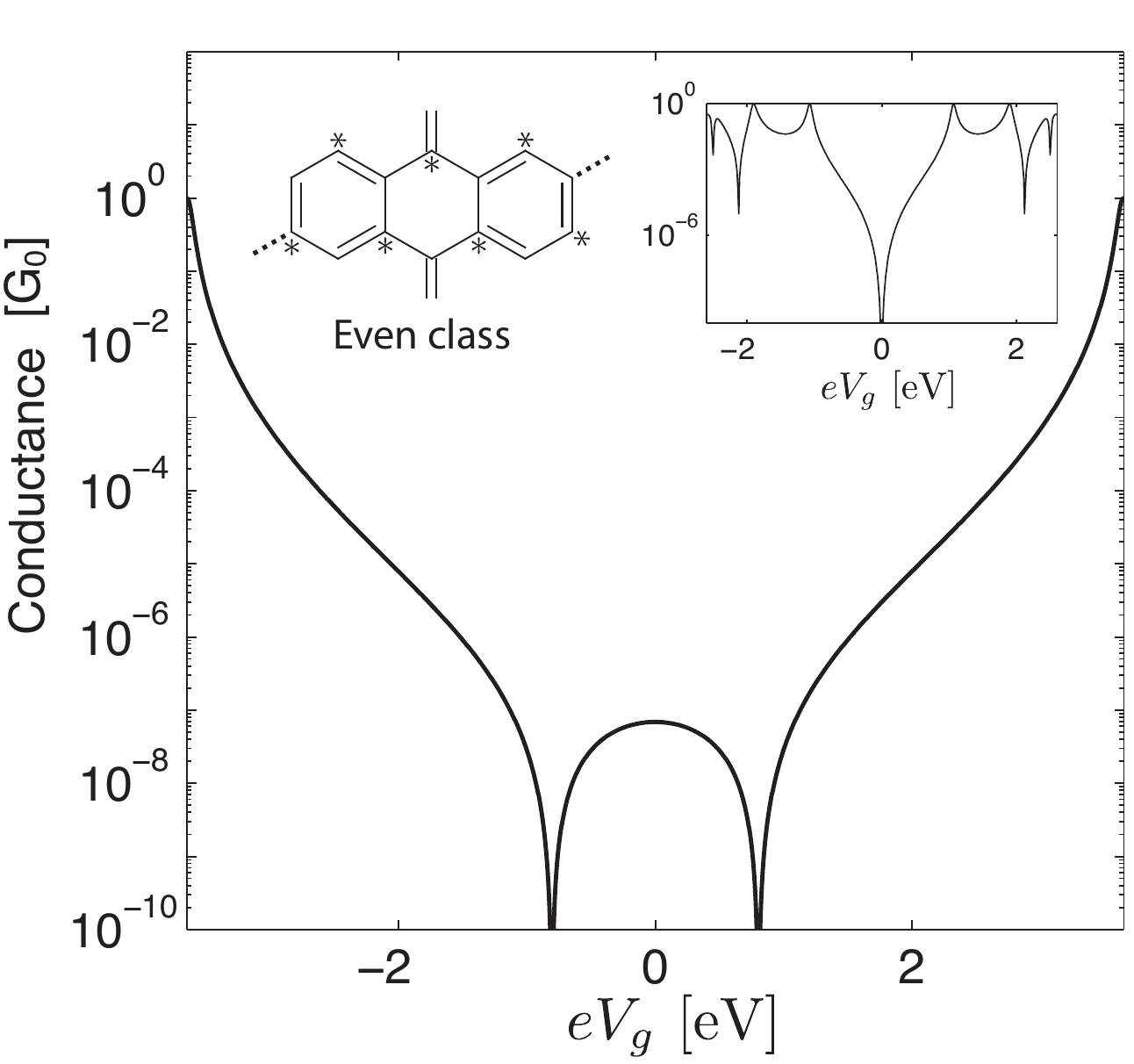}
  \caption{Off-resonant cotunneling conductance through the $\pi$-system of various conjugated alternant hydrocarbon molecular junctions. The conductance is calculated from Eq. \eqref{eq:Gzero} by exact diagonalization of the full PPP model including intra-molecular interactions and setting $\Gamma_s \Gamma_d  = 0.01 (\textrm{eV})^2 $. The dashed arms symbolize binding sites between the electrodes and the molecule. The calculated conductance conforms with the starring rule (summarized in the table). When entering and leaving through two sites belonging to different sublattices (star, no-star), the result is an even number of nodes. When the two connected sites belong to the same sublattice (e.g. star, star), there is an odd number of nodes. The inset shows the corresponding transmission for a simple H\"{u}ckel model.}
  \label{fig:table}
\end{figure*}

In general, this classification of interference in off-resonant quantum transport requires a detailed numerical calculation of the exact many-body eigenstates involved in the FD orbitals defining $\Qi$. For neutral alternant (bipartite) hydrocarbon $\pi$-systems, however, the classification can readily be carried out using a simple starring procedure. In an alternant system every other $p_z$-orbital can be marked by a star, such that all starred orbitals only have non-starred neighbors and vice versa. An equal number of starred and non-starred orbitals assures a spin-singlet ground state~\cite{Lieb1989} and the conductance formula~\eqref{eq:Gzero} is valid. Attaching the molecule to electrodes through $p_z$-orbitals on disjoint sub-lattices (star and non-star) makes the interference class \emph{even}, whereas contacting two $p_z$-orbitals on the same sub-lattice (both starred or both unstarred) makes it \emph{odd}.

This rule is a consequence of the Coulson-Rushbrooke-McLachlan pairing theorem~\cite{Coulson1940,McLachlan1959,McLachlan1961} for alternant hydrocarbon PPP models. For completeness we include a proof of the theorem in appendix \ref{sec:pairing}, establishing that for any $N$-electron eigenstate of the bipartite molecular Hamiltonian, represented by an extended Hubbard model, there is a corresponding $N$-hole eigenstate with the same energy. The proof is a straightforward application of the anti-unitary staggered particle-hole transformation, $\mathcal{U}$, defined by
\begin{equation}
\trans{ (z \, \create{c}{i \sigma} ) }= z^* \, (-1)^i \annihilate{c}{i \sigma}, \qquad \mbox{for $i = 1\ldots N_a$},
\end{equation}
where the orbital index, $i$, is chosen odd on starred sites and even on un-starred sites of the alternant molecule at hand. It is this transformation which transforms an eigenstate, $|\Psi^N_n\rangle$, with $N$ electrons into an eigenstate,
\begin{align}
|\Phi^{2N_a - N}_n\rangle=\mathcal{U}|\Psi^N_n\rangle,
\end{align}
with $2N_a - N$ electrons, i.e. $N$ holes, having the same energy.
Since all eigenstates are paired in this way, it is clear that any $N$-electron state $|\Psi_n^{N}\rangle$ represents the same state as some pairing partner $|\Phi_{m}^{N}\rangle = \mathcal{U} |\Psi^{2N_a - N}_m\rangle$. This implies that these two states can only differ by a trivial phase-factor, $\gamma^{N}_{mn}$, that is
\begin{align}
\mathcal{U}|\Psi^{2N_a - N}_m\rangle=|\Phi_m^{N}\rangle=e^{i\gamma^{N}_{mn}}|\Psi_n^{N}\rangle.
\end{align}
In other words, there is a one to one correspondence between the complete set of energy $N$-electron eigenstates $\{|\Psi_n^{N}\rangle\}$ and $\{|\Phi_m^{N}\rangle\}$, with the phase-factors, $e^{i\gamma^{N}_{mn}}$ defining the unitary transformation between the two.

Using this relation, we can now establish the following useful connection between FD-orbitals for respectively adding, or removing an electron from the molecule
\begin{align}
\langle\Psi_m^{2N_{a}-N+1}|&\create{c}{i\sigma}|\Psi_{m'}^{2N_{a}-N}\rangle\\
&=
\langle\Psi_m^{2N_{a}-N+1}|\mathcal{U}^\dagger\trans{\create{c}{i\sigma}}\mathcal{U}|\Psi_{m'}^{2N_{a}-N}\rangle
\nonumber \\
&=(-1)^{i}\langle\Psi_n^{N-1}|\annihilate{c}{i\sigma}|\Psi_{n'}^{N}\rangle
e^{i(\gamma^{N-1}_{mn}-\gamma^{N}_{m'n'})}.\nonumber 
\end{align}
In the last line we have used the anti-unitary nature~\footnote{An anti-unitary operator is not linear, and to alleviate this problem one can write the transformed state \mbox{$\mathcal{U} \ket{a} = \ket{\mathcal{U} a}$}. Then the anti-unitarity implies that \mbox{$\langle \mathcal{U} a | \mathcal{U} b \rangle = \langle b | a \rangle$}, which can be applied multiple times to show that \mbox{$\langle a | \hat{O} | b \rangle = \langle \mathcal{U} b | (\mathcal{U} \hat{O} \mathcal{U}^\dagger)^\dagger | \mathcal{U} a \rangle$}. This is the same result as for a unitary operator except for the complex conjugation.} of the symmetry transformation $\mathcal{U}$.

For a half-filled hydrocarbon, i.e.~\mbox{$N = N_a$}, with a non-degenerate spin-singlet ground state $|\Psi_0^{N_a}\rangle$, this implies the following simple relation between the $N_a$-electron ground state ato the $N_a+1$-electron ground state FD-orbitals:
\begin{align}
\langle\Psi_{0,m}^{N_{a}+1}|&\create{c}{i\sigma}|\Psi_{0}^{N_{a}}\rangle\\
&=
(-1)^{i}\langle \Psi_{0}^{N_{a}} |\create{c}{i\sigma}|\Psi_{0,n}^{N_{a}-1} \rangle
e^{i(\gamma^{N_{a}-1}_{mn}-\gamma^{N_{a}}_{00})},\nonumber 
\end{align}
where we explicitly take into account that the $N_{a}\pm 1$ electron ground states may be degenerate. If nothing else, then at least the spin $\sigma$ of the added/removed electron introduces such a degeneracy.
From this, we can now re-express $\Qi$ as follows:
\begin{align}
\Qi&=
\frac{
\sum_{n}
\bra{\Psi_{0}^{N_{a}}} \create{c}{i_s \sigma}\ket{\Psi_{0,n}^{N_{a}-1}}
\bra{\Psi_{0,n}^{N_{a}-1}}\annihilate{c}{i_d \sigma}\ket{\Psi_{0}^{N_{a}}}
}{
\sum_{m}
\bra{\Psi_{0}^{N_{a}}}\annihilate{c}{i_d \sigma} \ket{\Psi_{0,m}^{N_{a}+1}}
\bra{\Psi_{0,m}^{N_{a}+1}}\create{c}{i_s \sigma}\ket{\Psi_{0}^{N_{a}}}
}\nonumber\\
&=(-1)^{i_s}(-1)^{i_d}
e^{i(\gamma^{N_{a}}_{00}-\gamma^{N_{a}-1}_{mn'})}
e^{-i(\gamma^{N_{a}}_{00}-\gamma^{N_{a}-1}_{mn'})}
\nonumber \\
&\qquad
\times \frac{
\sum_{n}
\bra{\Psi_{0}^{N_{a}}} \create{c}{i_d \sigma}\ket{\Psi_{0,n}^{N_{a}-1}}
\bra{\Psi_{0,n}^{N_{a}-1}}\annihilate{c}{i_s \sigma}\ket{\Psi_{0}^{N_{a}}}
}{
\sum_{n'}
\bra{\Psi_{0}^{N_{a}}} \create{c}{i_d \sigma}\ket{\Psi_{0,n'}^{N_{a}-1}}
\bra{\Psi_{0,n'}^{N_{a}-1}}\annihilate{c}{i_s \sigma}\ket{\Psi_{0}^{N_{a}}}
}
\nonumber\\
&=(-1)^{i_{s}+i_{d}}, \label{QI1}
\end{align}
where the phase factors are seen to cancel.

This result is surprisingly simple, so let us reiterate its implications. When the two connecting orbitals $i_s$ and $i_d$ belong to the same sub-lattice (starred or unstarred), we are in the odd class ($\Qi > 0$), and when they belong to disjoint sub-lattices (one starred, one unstarred), we are in the even class ($\Qi < 0$). Tracing a path through the molecule between the connecting orbitals, the interference class is also given by the number of atoms visited by the path. An odd number of atoms imply that the transport is characterized by the odd interference class, while an even number of atoms implies the even interference class.

This straight-forward starring rule is one of the main results of the present paper. Including the full effects of intra-molecular interactions we numerically calculate the cotunneling conductance for various alternant hydrocarbon molecular junctions by exact digitalization of the corresponding PPP model. The results are shown in Fig.~\ref{fig:table}, where the starring rule has been summarized in a small table. The reader is invited to try his hand at the starring rule, and confirm that the rule correctly predicts the number of interference nodes.
In Fig.~\ref{fig:table} it is also shown how the interference nodes of non-interacting H\"{u}ckel models of neutral alternant hydrocarbons are often degenerate. In the case of double node degeneracy the classification directly shows which nodes will split (or lift) when including intra-molecular interactions. Also note that this classification ensures the presence of an interference node for molecular junctions in the odd class regardless of the strength of the interactions. Hence both the $\Qi$ classification and the starring rule represent useful and simple tools. This is true even when working with interference in H\"{u}ckel models where \citet{Markussen2010} have derived a set of graphical rules. Note that when neglecting electron-electron interactions on the molecule both the Markussen rules and the starring rule apply and are consistent in their even-odd classification.



\subsection{Interference in Transport Through Stilbene}

\begin{figure*}[tb]
\includegraphics[width=1\textwidth]{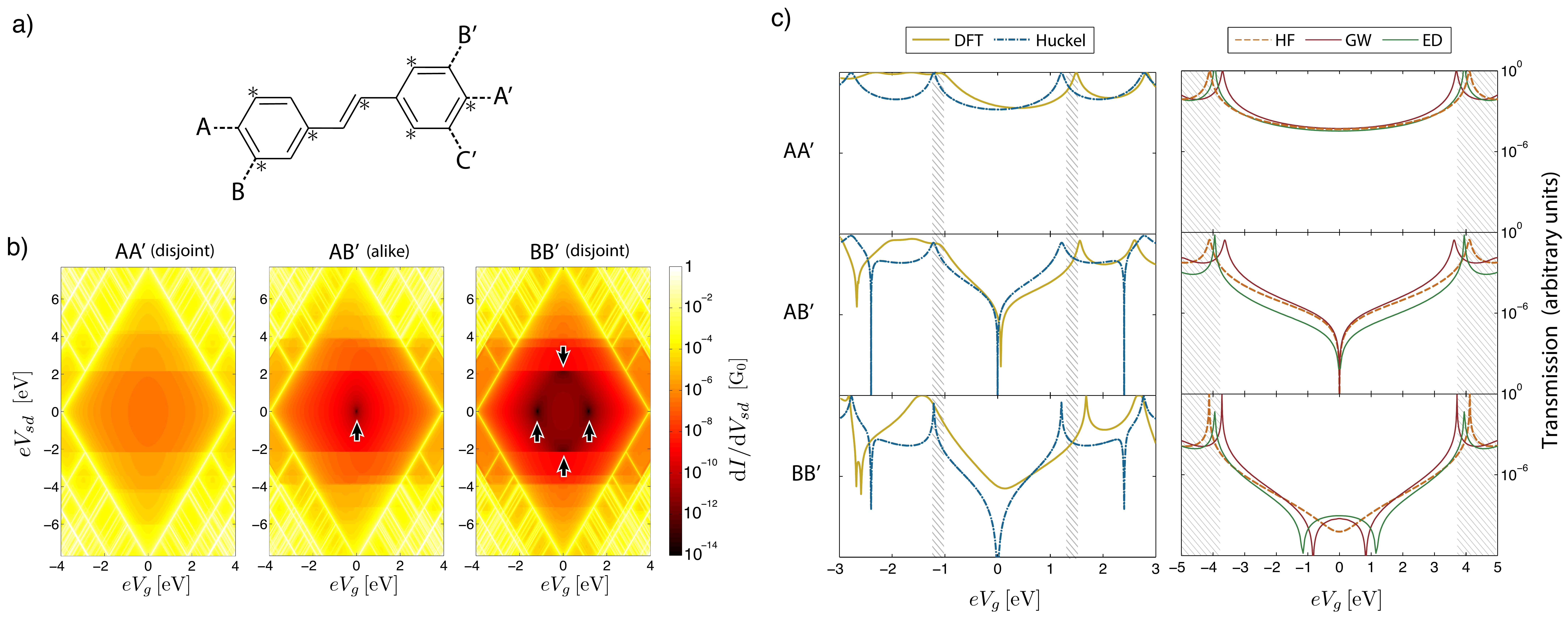}
\caption{(color online) The stilbene molecular junction. \textbf{a)} The molecular junction setup. \textbf{b)} The stilbene molecule. The dashed arms symbolize possible binding sites between the electrodes and the molecule. The stars refer to the quantum interference classification explained in Figure~\ref{fig:qi}. \textbf{c)} Stilbene stability diagrams for the neutral $\pi$-system for $\Gamma_s \Gamma_d  = 0.01 (\textrm{eV})^2 $ in different coupling configurations obtained from the current expression in equation \eqref{eq:singlet_current}. In the AB' and BB' configurations all the interference nodes in the conductance are marked with arrows. Note that in the BB' configuration the two nodes at non-zero bias are hidden due to inelastic cotunnelling processes. \textbf{d)} Transmission through the $\pi$-system of substituted stilbene calculated by DFT, HT, HF, GW and ED, respectively. The edge of the shaded region marks the charge degeneracy points, while the unshaded regions denote the gate values of the relevant charge state. The mean-field methods (DFT and HF) and the non-interacting HT all predict interference, but fail to capture the double node of the even interference class predicted by the starring rule for the BB' configuration.}
\label{fig:stilbene_transmission}
\end{figure*}

As an example, we now provide a detailed analysis of the stilbene molecule shown in Fig.~\ref{fig:stilbene_transmission}(a). Being an alternant hydrocarbon, we can employ the starring rule described above, and we immediately conclude that this molecule will have an even number of conductance nodes when contacted in AA' or BB' configuration, and an odd number of nodes in the AB' configuration. Within a PPP-model description of the $\pi$-system, this molecule is still amenable to exact diagonalization. We show the stability diagrams for the different contacting geometries in Fig.~\ref{fig:stilbene_transmission}(b), showing the differential conductance $d I/dV_{sd}$ on a logarithmic scale as a function of a backgate voltage, $V_g$, and source drain voltage, $V_{sd}$, calculated at zero temperature. The bright colors indicate a high differential conductance with the inner diamond demarcating the (red-black) regions, inside which our off-resonant current formula for elastic cotunneling becomes valid. The dark spots in the middle and right most panels in Fig.~\ref{fig:stilbene_transmission}(b) show unusual, strong suppressions of the differential conductance, related to the destructive interference in the cotunneling conductance. Note also how inelastic cotunneling processes become relevant at certain values of $V_{sd}$ and instantly cuts off any interference features present in the elastic cotunneling current. We have confirmed that the salient features of Fig.~\ref{fig:stilbene_transmission}(b), are indeed reproduced for a simple quinoid type molecule by a full generalized master equation calculation~\cite{Leijnse2008,Koller2010} (not shown).

Fig.~\ref{fig:stilbene_transmission}(c) shows the zero-bias conductance in three different contacting geometries, and calculated using respectively DFT or HT, or HF, GW or ED. More details on the calculations are given in Appendix \ref{app:dft}. Whereas all methods agree on the interference class of AA' and AB' configuration, a clear disagreement arises in BB' configuration, where HT shows only one node and DFT and HF both predict no nodes or at best a single local minimum. Only GW and exact diagonalization agree on two nodes, consistent with our simple starring procedure. Interpreting the DFT and HF results as predicting no nodes, and HT as predicting two degenerate nodes, they may all be said to give the correct class, but comparing with GW and ED, it is clear that the correct result depends crucially on a careful consideration of Coulomb interactions, as noted also in Refs.~\nocite{Solomon2011,Barr2013}\citenum{Solomon2011,Barr2013}. Calculating $\Qi$ within HT correctly predicts an even interference class, and the degenerate node is an accidental degeneracy, that will be lifted by interactions as found by GW and ED, whereas DFT and HF both remove the node altogether. We note that the DFT results can depend on the functional being used, and it is conceivable that all three possibilities (no node, degenerate nodes or split nodes) can be observed with different functionals. As illustrated by this example, our classification scheme serves as a valuable tool for settling such ambiguities.


\section{Transport through a spin-doublet ground state}

The previous analysis may be extended to molecules with \textit{degenerate} ground states, e.g. $\pi$-systems with an odd number of electrons. For such systems the ground state is usually a spin doublet $\ket{\Psi^{N}_{0,m}}$, here denoted by the spin index $m=\,\,\uparrow,\downarrow$. Combining particle and hole amplitudes into the transport amplitude
\begin{align}
A_{m,l}^{\gamma \sigma }(V_{g}) = h_{m,l}^{\gamma \sigma}(V_g,0) + p_{m,l}^{\gamma \sigma} (V_g,0),
\end{align}
the zero-bias conductance can be written as the sum of three different terms,
\begin{align}
  G &= \frac{1}{2}\frac{e^2}{h} \Gamma_s \Gamma_d \sum_{\sigma} \left(
    |A_{\sigma,\sigma}^{\sigma \sigma}|^2
    + |A_{\sigma,\sigma}^{\bar{\sigma} \bar{\sigma}}|^2
    + |A_{\bar{\sigma},\sigma}^{\bar{\sigma} \sigma}|^2
    \right),
  \label{eq:Gdoublet0}
\end{align}
where $\bar{\sigma}$ denotes the opposite of $\sigma$, i.e. $\bar{\uparrow} = \downarrow$ and vice versa.
Using the spin-rotation symmetry of the Hamiltonian, a bit of algebra shows that for doublet ground states
\begin{align}
A_{m,m}^{\uparrow \uparrow} - A_{m,m}^{\downarrow \downarrow} = \pm A_{\downarrow,\uparrow}^{\downarrow \uparrow}, \hspace*{4mm} {\rm for}\hspace*{2mm} m = \uparrow/\downarrow.
\end{align}
We can then define the potential scattering amplitude
\begin{align}
W = \sum_{\sigma} A^{\sigma \sigma}_{\uparrow,\uparrow} = \sum_{\sigma} A^{\sigma \sigma}_{\downarrow,\downarrow},
\end{align}
and the exchange amplitude
\begin{align}
J = A^{\uparrow \downarrow}_{\uparrow, \downarrow} = A^{\downarrow \uparrow}_{\downarrow, \uparrow}.
\end{align}
in terms of which, the zero-bias conductance becomes
\begin{align}\label{eq:Gdoublet}
G
&=G_W + G_J \\
&=\frac{e^2}{2 h}\Gamma_s\Gamma_d\left(|W|^2+3|J|^2\right).\nonumber
\end{align}
Clearly an interference dip in the conductance demands coincident dips in both $W$ and $J$. In the following we show that this is generally not possible, since the requirement for interference in the two are mutually exclusive. Notice that Eq.~\ref{eq:Gdoublet} corresponds exactly to the conductance for an effective Kondo model, including both potential scattering and exchange tunneling terms~\footnote{Compare e.g. to  Eq. (10.96) in Ref.~\onlinecite{BruusFlensberg}, in which $W$ and $J$ were defined slightly differently and an erroneous factor of 1/2 appears on the potential scattering term}.

\subsection{Interference Classification of Transport through Alternant Hydrocarbons}

Here we again restrict the analysis to neutral alternant hydrocarbons but now with an odd number of orbitals in the $\pi$-system, in order to allow for a spin-doublet ground state. The Rushbrooke-Coulson-McLachlan pairing theorem again allows us to derive a simple starring rule, from which we may classify interference dips in the two cotunneling amplitudes, $W$ and $J$.

Since the classification of the potential scattering $\Qi_W = (-1)^{i_s + i_d}$ is similar to the previously discussed spin-singlet case, we here focus on deriving the classification of the exchange term, which is generally given as:
\begin{equation}\label{QS}
  \Qi_J =
    \frac{\bra{\Psi_{0,\bar{\sigma}}^{N_{a}}} \create{c}{i_s \bar{\sigma}} \ket{\Psi_0^{N_{a}-1}} \bra{\Psi_0^{N_{a}-1}}\annihilate{c}{i_d \sigma}\ket{\Psi_{0,\sigma}^{N_{a}}}
    }{
      \bra{\Psi_{0,\bar{\sigma}}^{N_{a}}}\annihilate{c}{i_d \sigma} \ket{\Psi_0^{N_{a}+1}} \bra{\Psi_0^{N_{a}+1}}\create{c}{i_s \bar{\sigma}}\ket{\Psi_{0,\sigma}^{N_{a}}}
    },
\end{equation}
where we again imply a sum over the possible degeneracy of the $N_{a}\pm1$ states.
As demonstrated by formula~\eqref{transS} in Appendix A, the staggered particle-hole transformation $\mathcal{U}$ used in the pairing theorem ensures that each eigenstate has a symmetry partner with the opposite spin. For the neutral spin-doublet ground state that is
\begin{align}
\mathcal{U} |\Psi_{0,\sigma}^{N_a}\rangle = e^{i\gamma^{N_{a}}_{\sigma\bar{\sigma}}} |\Psi_{0,\bar{\sigma}}^{N_a}\rangle, \qquad \mbox{for}\,\, \sigma = \uparrow/\downarrow.
\end{align}
These states are of course also related by a spin reversal, 
\begin{equation}\label{rotatePsi}
\hat{R}_\pi |\Psi_{0,\sigma}^{N_a}\rangle = i |\Psi_{0,\bar{\sigma}}^{N_a}\rangle,
\end{equation}
where $\hat{R}_\pi = \exp(i \pi \hat{S}_x)$, which transforms under the anti-unitary
symmetry transformation $\mathcal{U}$ as
\begin{equation}\label{TransRx}
\trans{\hat{R}_\pi} = \trans{e^{i\pi S_x}} = e^{-i\pi S_x}=\hat{R}_\pi.
\end{equation}
From this, a transformation of Eq.~\eqref{rotatePsi} readily shows that
\begin{align}
i e^{i\gamma^{N_{a}}_{\bar{\sigma}\sigma}}|\Psi_{0,\sigma}^{N_a}\rangle = -i e^{i\gamma^{N_{a}}_{\sigma\bar{\sigma}}} |\Psi_{0,\sigma}^{N_a}\rangle,
\end{align}
which in turn implies the following relation between the doublet phases:
\begin{align}\label{Jphas}
\gamma^{N_{a}}_{\sigma\bar{\sigma}}=\gamma^{N_{a}}_{\bar{\sigma}\sigma} + \pi.
\end{align}

As for the non-degenerate case, we can again use the pairing theorem to rewrite the FD-orbital. With the spin-doublet index on the ground state, one now finds
\begin{align}
\langle\Psi_{0,m}^{N_{a}+1}|&\create{c}{i_{s}\bar{\sigma}}|\Psi_{0,\sigma}^{N_{a}}\rangle\\
&=
(-1)^{i}\langle \Psi_{0,\bar{\sigma}}^{N_{a}} |\create{c}{i_{s}\sigma}| \Psi_{0,n}^{N_{a}-1} \rangle
e^{i(\gamma^{N_{a}}_{\sigma\bar{\sigma}}-\gamma^{N_{a}-1}_{mn})},\nonumber 
\end{align}
and similarly for $\langle \Psi_{0,\bar{\sigma}}^{N_{a}} | \annihilate{c}{i_d \sigma} | \Psi_{0,m}^{N_{a}+1} \rangle$, which allows the following simple rewriting of Eq.~\eqref{QS}:
\begin{equation}\label{QS1}
  \Qi_J = - (-1)^{i_s+i_d}.
\end{equation}
As for the non-degenerate case, the phase factors of the virtual intermediate states cancel, whereas the phases coming from the transformation of the initial, and final $N_{a}$-electron states, which now have opposite spin, combine to an extra factor of $-1$ due to Eq.~\eqref{Jphas}.
This has the important consequence that $\Qi_J = - \Qi_W$ for alternant lattices, implying that the two distinct contributions to the conductance always belong to \emph{different} interference classes.

This mutual exclusion of interference, $\Qi_J = - \Qi_W$, is actually more general than this statement for a half-filled molecule, and can be shown to hold for all spin-doublet ground states $\ket{\Psi_{0,m}^N}$ as long as the neighboring charge states, $N\pm1$, both have spin-singlet ground states~\cite{Pedersen2013}.

\subsection{Interference in Transport Through a Biphenyl Molecule}

\begin{figure*}[t]
    \centering
    \includegraphics[width=.7\columnwidth]{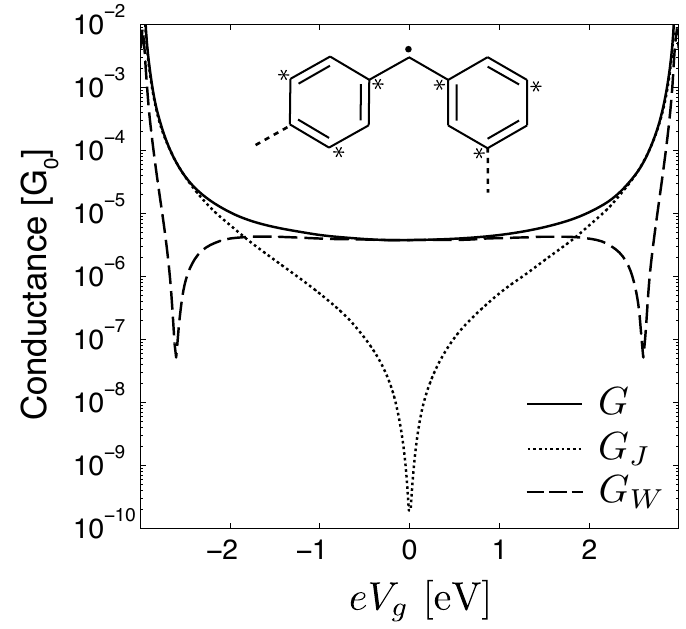}
    \hspace{1.5cm}
    \includegraphics[width=.7\columnwidth]{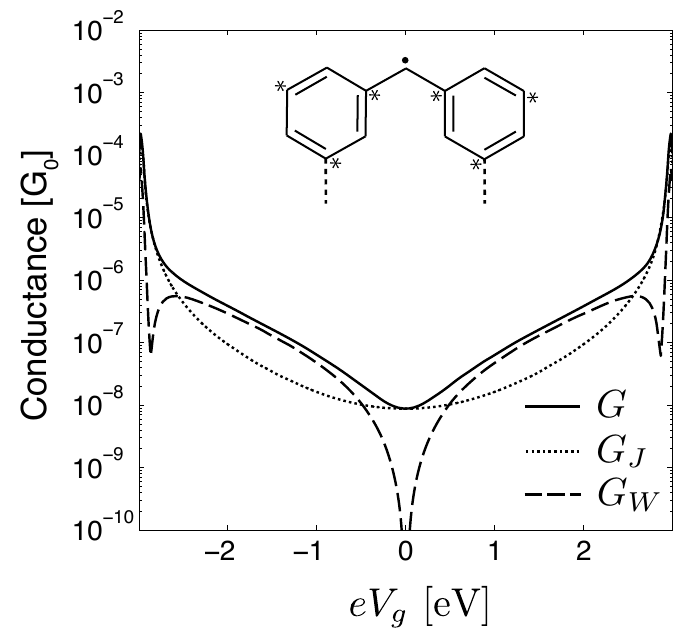}
  \caption{The biphenyl molecular junction with a spin-doublet ground state. Plot of the zero-bias conductance of a biphenyl molecular junction in two different configurations using $\Gamma_s \Gamma_d  = 0.01 (\textrm{eV})^2 $. The dashed arms symbolize possible binding sites between the electrodes and the molecule. Note how the interference dips appear at different positions for the $W$ and $J$ channel, and therefore do not show up in the total current. The configuration in the right panel shows how  the exchange term  fills in the dip in the potential scattering amplitude, leaving only a shallow local minimum near $V_{g}=0$.}
  \label{fig:biphenylx}
\end{figure*}

As an example, we show the calculated linear conductance for a biphenyl molecule with an odd number of electrons in the neutral $\pi$-system in Fig.~\ref{fig:biphenylx}. This is an odd-alternant hydrocarbon, and our starring rules readily show that $\Qi_J = - \Qi_W = -(-1)^{i_s+i_d}$, which takes the values $1$ and $-1$, for entry and exit sites belonging to different or same sublattice, respectively.
That is, entering and leaving the molecule on alike or disjoint (starred/unstarred) sublattice sites decides whether $W$ or $J$ has a zero near the middle of the charge state (see Fig.~\ref{fig:biphenylx}). We note that this is at odds with the interference dip suggested in Ref.~\nocite{Bergfield2011}\citenum{Bergfield2011} to appear already within a single-orbital Anderson model, and it seems that the exchange term, $J$, has been missed there.

In a single-orbital Anderson model it is always the potential scattering term, $W$, which vanishes at the particle-hole symmetric point~\cite{BruusFlensberg}. This is consistent with our starring rule, since that single molecular orbital will necessarily act as both entry and exit point for the electrons tunneling to and from the leads, whereby $\Qi_J = - \Qi_W=-1$ corresponding to the odd interference class for $W$ alone. With a molecule allowing for two distinct entry and exit points, however, we see that the situation can also be reversed, so as to observe a node in $J$, while $W$ remains finite. This is shown for the BB'-contacted biphenyl in the right panel of Fig.~\ref{fig:biphenylx}.

A molecule off resonance with the leads and with a spin degenerate ground state is known to exhibit Kondo effect, manifest as a sudden increase of the zero-bias conductance, when temperature is lowered past the a characteristic Kondo temperature, $T_{K}$~\cite{BruusFlensberg}. In general, the zero-bias conductance peak depends not only on inter-lead cotunneling, encoded in $J$, but also on repeated intra-lead cotunneling, and we shall defer a more thorough analysis of the intricate interplay of these processes with interference in a separate publication. At any rate it is clear that a sharp dip in $J$ when varying the gate-voltage across the relevant charge state, should lead to marked variations in the Kondo conductance peak itself, whenever the molecule is contacted via different sublattices, i.e. at one starred, and one unstarred site. Unusual gate-dependences of the Kondo conductance peak have indeed been reported to appear in a number of different single-molecule devices~\cite{Natelson2008,Scott2010,Zyazin2010,Fock2012}, and with the simple rules offered in this paper, one might revisit these cases to assess any possible links to a likely contacting geometry. Depending on the exact contacting geometry, these molecular junctions could have multiple entry and/or exit points, which will of course complicate the analysis considerably. Already from the study of simple two, or three orbital Anderson models with multiple orbitals connected to each lead, the appearance of transmission nodes is known to have a rather non-trivial dependence on the various parameters~\cite{Karrasch2007,Hecht2009}. Therefore it is only in cases with a clearly dominating pair of contacting sites, that our rules apply in this straightforward manner. Our approach can of course be readily generalized to deal with multiple contacting sites, but clearly the interference problem at quickly becomes rather complicated and requires additional information about the relevant parameters.


\section{Conclusion}

In conclusion, we have put forth a simple set of rules for classifying interacting off-resonant single-molecule junctions into two distinct interference classes with either an even or an odd number of transmission minima. For alternant hydrocarbons, this classification reduces to a simple starring rule, with an extension to the spin-degenerate case, which revealed mutually exclusive interference dips in the competing potential scattering and exchange amplitudes. This provides a powerful tool to discriminate between different numerical results, for which more reliable calculations are not readily available.

Note that molecular vibrations have been neglected in this work, and indeed a sufficiently weak electron-phonon coupling should have little or no influence on the interference aspects described here, except for new inelastic cotunneling channels showing up at finite bias (cf. Fig. 2c). For stronger electron phonon coupling, however, one might expect an interesting interplay between interference nodes and Franck-Condon blockade~\cite{Koch2006} which deserves closer investigation. Both for this purpose and to investigate closer the details of Kondo screening with interference nodes in $J$, it should be instructive to study the strong coupling regime (beyond leading order in $\Gamma{s,d}/\varepsilon_{0}^{p,h}$), possibly by means of numerical, or functional renormalization group (RG) techniques. Already at the level of perturbative RG, a number of interesting conclusions can be drawn on Kondo effects in single-molecule junctions. These issues are beyond the scope of the present paper and will be deferred to a separate publication.

The research leading to these results has received funding from the European Union Seventh Framework Programme (FP7/2007-2013) under agreement no 270369 (ELFOS) and from the European Research Council, ERC Grant agreement no 258806. M. Leijnse acknowledges financial support from the Swedish Research Council (VR). The Center for Quantum Devices is funded by the Danish National Research Foundation

\appendix

\section{The Pairing Theorem for Alternant Hydrocarbons}\label{sec:pairing}

In general we describe the $\pi$-electrons on the molecule by the extended Hubbard (or Pariser-Parr-Pople) model, which we split up into components,
\begin{equation}\label{eq:Htot}
  H = \hat{H}_T + \hat{H}_\varepsilon+ \hat{H}_U + \hat{H}_V.
\end{equation}
Here
\begin{align}
  \hat{H}_T &= \sum_{ij,\sigma} t_{ij} \create{c}{i \sigma} \annihilate{c}{j \sigma} +h.c.\\
  \hat{H}_\varepsilon &= \sum_i \varepsilon_i \n{i} \\
  \hat{H}_U &= \sum_i U_i (\n{i \uparrow} - \tfrac{1}{2}) (\n{i \downarrow} - \tfrac{1}{2})\\
  \hat{H}_V &= \sum_{i\neq j} V_{ij}(\n{i}-1)(\n{j}-1),
\end{align}
with $\n{i \sigma}=\create{c}{i \sigma} \annihilate{c}{i \sigma}$, and $\n{i}=\n{i\uparrow}+\n{i\downarrow}$. The molecule consists of $N_a$ atoms, each with a $p_z$ orbital, hence $\create{c}{i\sigma}$ creates an electron with spin $\sigma$ in the $p_z$ orbital of atom number $i$.

The extended Hubbard Hamiltonian has a particular symmetry when the molecule can be divided into two subsets, $A$ and $B$, of atoms, such that the hopping amplitudes only connect atoms which belong to different subsets, i.e.~$t_{ij} = 0$, if $i,j$ both are in $A$ or both in $B$. Such a molecule is called {\em alternant} or {\em bipartite}. To understand this symmetry, consider the anti-unitary transformation, which transforms complex numbers, $z$, and creation operators as follows
\begin{equation}\label{transDef}
\trans{ (z \, \create{c}{i \sigma} ) }= z^* \, (-1)^i \annihilate{c}{i \sigma}, \qquad \mbox{for $i = 1\ldots N_a$}.
\end{equation}
Here we use the notation
\begin{equation}\label{m1}
  (-1)^i = \left\{ \begin{array}{ll} 1 & \mbox{$i$ in $A$}\\ -1 & \mbox{$i$ in $B$}\end{array}\right.
\end{equation}
The transformation $\mathcal{U}$ is then composed of a particle-hole transformation and an overall sign $(-1)^i$ followed by complex conjugation.

The number operators thus transform as \mbox{$\trans{\n{i\sigma}} = 1-\n{i\sigma}$}, while the various terms in the Hamiltonian transform as
\begin{align}
\trans{\hat{H}_T} &= \sum_{ij,\sigma} (-1)^i (-1)^j  \annihilate{c}{i\sigma}\create{c}{j\sigma} = \hat{H}_T, \label{eq:HTpairing} \\
\trans{\hat{H}_\varepsilon} &= \sum_i\varepsilon_i (2-\n{i}), \\
\trans{\hat{H}_U} &= U \sum_i (\tfrac{1}{2}-\n{i\uparrow})(\tfrac{1}{2}-\n{i\downarrow}) = \hat{H}_U,  \\
\trans{\hat{H}_V} &= \sum_{i\ne j} V_{ij}(1-\n{i})(1-\n{j}) = \hat{H}_V.
\end{align}
Note that the anti-unitary pairing symmetry holds even when including complex hopping amplitudes through e.g. the spin-orbit interaction. 

The invariance of $\hat{H}_T$ relies heavily on the fact that the molecule is alternant. The term $\hat{H}_\varepsilon$ is not invariant. If, however, the energies $\varepsilon_i$ are independent of $i$, i.e.~all being equal to a fixed energy $\varepsilon_0$, then the transformation is
\begin{equation}
  \trans{\hat{H}_\varepsilon} = \hat{H}_\varepsilon + 2\varepsilon_0(N_a-\hat{N}_e), \text{ with }\hat{N}_e = \sum_i \n{i}.
\end{equation}
The full Hamiltonian now transforms as
\begin{equation}\label{transH}
  \trans{\hat{H}} = \hat{H} + 2\varepsilon_0 (N_a- \hat{N}_e).
\end{equation}
If we finally choose the zero of the one-electron energies such that $\varepsilon_0 = 0$, then the Hamiltonian is invariant under the transformation (\ref{transDef}).

We have specified how operators transform. We also need to specify how states in the Hilbert space transform. We only need to do that for the vacuum state $|0\rangle$ and for the completely filled state $|2N_a\rangle = \Pi_{i\sigma} \create{c}{i\sigma} |0\rangle$, since all other states can be obtained from these by application of creation operators and by use of the superposition principle. The states transform as follows
\begin{equation}\label{transV}
  \mathcal{U}|0\rangle = |2N_a\rangle, \qquad\mathcal{U}|2N_a\rangle = |0\rangle.
\end{equation}
It is easy to show that the last equation is consistent with the first and with the transformation rules for the creation operators.

For later use we also introduce the total spin operator. The spin operator is written in second quantized form
\begin{equation}\label{sTotal}
  \vS = \frac{\hbar}{2}\sum_{i\sigma\sigma'} \create{c}{i\sigma}\vPauli_{\sigma\sigma'}\annihilate{c}{i\sigma'}.
\end{equation}
This operator transforms as follows (here the anti-unitarity of $\mathcal{U}$ is important),
\begin{equation}\label{transS}
  \trans{\vS} = \frac{\hbar}{2}\sum_{i\sigma\sigma'}\annihilate{c}{i\sigma}\vPauli^*_{\sigma\sigma'}\create{c}{i\sigma'} = -\vS.
\end{equation}

Let us now consider a state with $N$ electrons, $|\Psi^{N}\rangle$. The transformed state $\mathcal{U}|\Psi^{N}\rangle$ will then be a state with $2N_a-N$ electrons. Let us in particular consider an eigenstate of the Hamiltonian:
\begin{equation}\label{eigen}
  \hat{H} |\Psi_n^{N}\rangle = E_n^N|\Psi_n^{N}\rangle.
\end{equation}
We denote the transformed state
\begin{equation}\label{transeigen}
  |\Phi_n^{2N_a-N}\rangle = \mathcal{U}|\Psi_n^{N}\rangle.
\end{equation}
Since the Hamiltonian is invariant under $\mathcal{U}$, this state is also an eigenstate of $\hat{H}$ with the same eigenvalues
\begin{equation}\label{transeigen1}
  \hat{H}|\Phi_n^{2N_a-N}\rangle = E_n^N |\Phi_n^{2N_a-N}\rangle.
\end{equation}
In this manner, the staggered particle-hole symmetry encoded in $\mathcal{U}$ implies a pairing between degenerate $N$-electron, and $N$-hole states, thus establishing the {\it pairing theorem} used in the main text.

Since the total spin-operator $\vS^2$ commutes with $\hat{H}$, we can classify the eigenstates by spin-quantum numbers, $S$ and $M$. An energy level with spin $S$ will be $2S+1$ times degenerate, and so will the transformed level. Furthermore, if the state $|\Psi_n^{(N)}\rangle$ is an eigenstate of $\hat{S}_z$ with eigenvalue $M$, then the transformed state will be an eigenstate of $\hat{S}_z$ with eigenvalue $-M$.

\section{DFT and GW Calculations} \label{app:dft}

We use DFT as implemented in the GPAW code to provide a quantum chemical description of charge transport through molecular junction systems~\cite{GPAW2010}. Molecules were optimized in the gas phase using the PBE exchange correlation (xc) functional~\cite{Perdew1996}. For all calculations the molecules were attached to the FCC hollow site of Au(111) with a Au-S bond length of $2.5$ \AA~(1.83~{\AA} above the surface). The scattering region supercell was modeled using 3-4 atomic Au layers on both sides of the molecule with $4\times4$ surface layer atoms. Periodic boundary conditions where used in the transverse directions and the 2D Brillouin zone was sampled using $4\times4$ $k$-points. We use semi-infinite atomistic leads and a double zeta polarized (DZP) basis set for all atoms in both the lead and the scattering region. Au atoms were frozen in the bulk lattice structure using the DFT derived lattice constant of $a=4.18$~\AA.

\begin{figure}[htb]
  \begin{center}
  \includegraphics[width=\columnwidth]{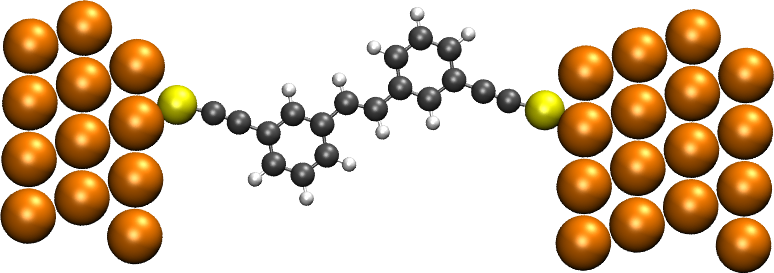}
  \caption{(color online) Relaxed geometry of substituted stilbene in the AA' configuration used for DFT calculations.}
  \label{fig:stilb_geom}
  \end{center}
\end{figure}

The transmission is calculated as a function of the energy $E-E_F$ at zero bias voltage and zero gate~\cite{Chen2012}. Here $E_F$ is the Fermi energy of the electrons in the leads. We assume that this energy-resolved transmission is similar to the transmission as a function of gate voltage as was observed in Ref.~\citenum{Chen2012}.

For calculations based on GW and HF we use the same semi-empirical Pariser-Parr-Pople (PPP) model Hamiltonian as for the exact-diagonalization (ED) results in the main text. The GW transport method is described in more detail in Refs. \citep{Thygesen2008,Strange2011,Strange2012}. Briefly, the Green's function of the contacted molecule is calculated from
\begin{equation}
  G_{ij}(\omega) = \left(
    \omega-H_0-\Sigma_{L}(\omega)-\Sigma_{R}(\omega)-\Sigma_\mathrm{xc}(\omega)
  \right)^{-1}_{ij},
\end{equation}

where $H_0$ is the noninteracting part of the Hamiltonian including the Hartree potential. $\Sigma_\alpha$ is the embedding self-energy from lead $\alpha$ accounting for the coupling to the semi-infinite leads. The semi-infinite leads are described using a nearest neighbor tight-binding chain with a large hopping element $t_L=-20$~eV, i.e. a featureless wide band lead. The last site of the semi-infinite tight-binding chains are coupled to the source $s$ and drain $d$ sites on the molecule, using a hopping element of $t_{s/d}=-\sqrt{|t_L|/2}$. This results in a broadening of the $s$ and $d$ sites. 

The self-energy is evaluated either using the Hartree-Fock or the $GW$ approximation. We evaluate the $GW$ self-energy fully self-consistently ensuring that conservation laws, such as charge conservation, are fulfilled. The energy dependence of $G$ and $\Sigma_\mathrm{xc}$ is sampled on a uniform grid $\omega_n = \varepsilon_n+i\eta$, where $\eta=0.001$~eV is an infinitesimal and $\varepsilon_n$ ranges from -100 to 100 eV, with a spacing of $\eta/2=0.0005$~eV.

\begin{figure}[htbp]
  \begin{center}
  \includegraphics[width=.8\columnwidth]{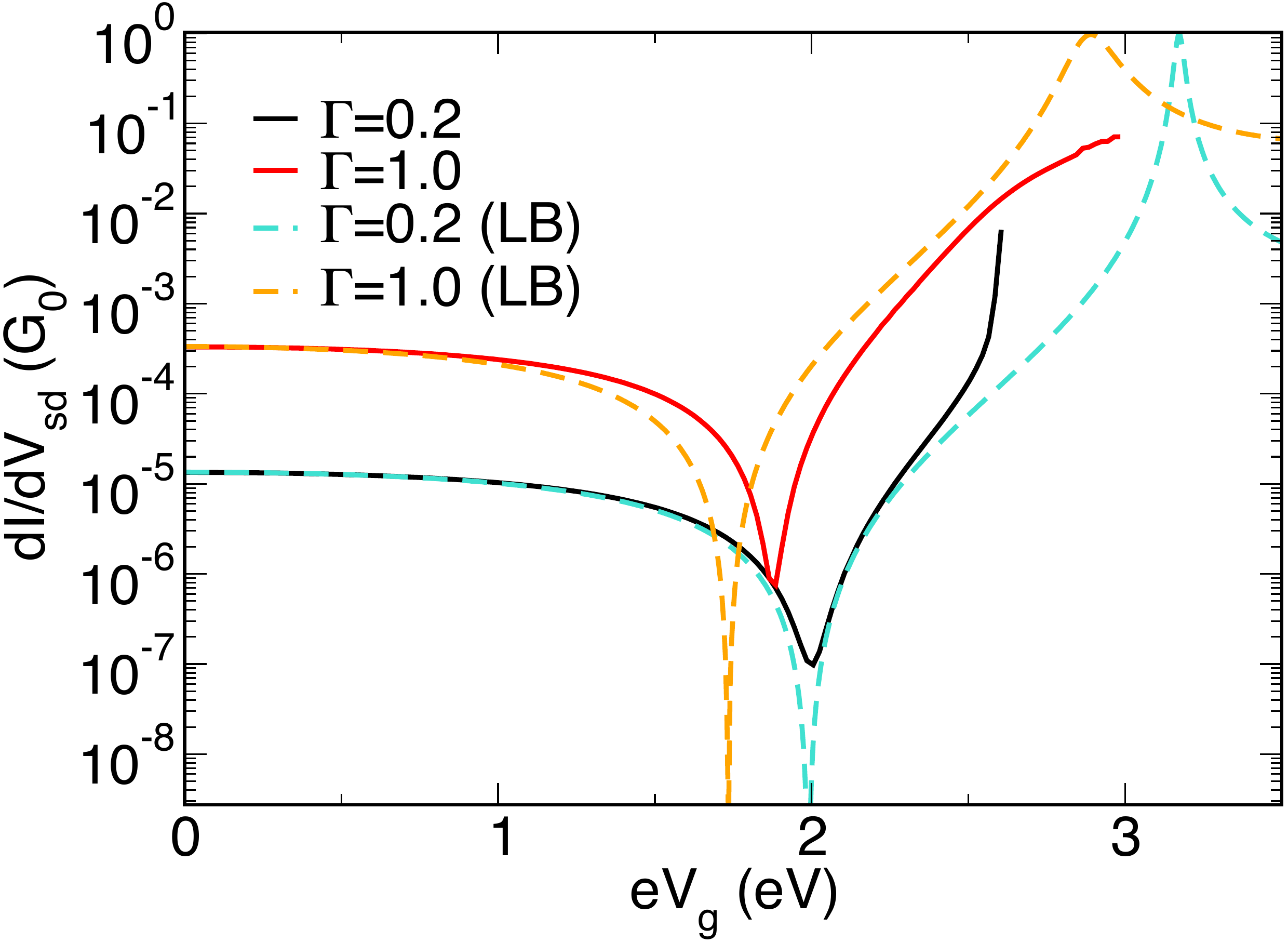}
  \caption{(color online) Low bias differential conductance as function of gate potential for the benzo-quinoid structure. The dashed and solid lines denotes the LB and Meir-Wingreen results, respectively.}
  \label{fig:GW_comparison}
  \end{center}
\end{figure}

Transport properties are for the DFT and GW methods calculated using the Landauer B{\"u}ttiker (LB) transmission formula expressed in terms of Green's functions
\begin{align}
  \tau(\varepsilon)=\text{Tr}\left[G^r(\varepsilon)\Gamma^L(\varepsilon) G^a(\varepsilon) \Gamma^R(\varepsilon)\right],
\end{align}
where $\Gamma^\alpha = i(\Sigma^r_{\alpha} - \Sigma^a_{\alpha})$ is given in terms of the lead $\alpha$ self-energy $\Sigma_\alpha$. We calculate the conductance in the zero bias voltage limit as $G=G_0\int (-n_F'(\varepsilon,T)) \tau(\varepsilon) d\varepsilon $, where $n_F'(\varepsilon,T)$ is the derivative of the Fermi function with respect to energy. $T$ is the temperature and $G_0=2e^2/h$ is the quantum of conductance, where $h$ and $e$ is Planck's constant and the electronic charge, respectively. For the benzene-quinoid structure, we have verified that a calculation of the low bias $dI/dV$ as a function of a gate voltage $V_g$ in the GW approximation, where the current is obtained using the Meir-Wingreen formula\cite{Meir1992}, gives the same destructive interference features as using the LB transmission function expression.



\bibliography{references}

\end{document}